\documentclass[11pt]{article}
\usepackage{graphicx}
\usepackage{xcolor}
\usepackage{amssymb}
\usepackage{amsmath}
\usepackage{mathabx}
\usepackage{bbm}
\usepackage{multirow}
\usepackage{multicol}
\usepackage{amscd}
\usepackage{mfpic}
\usepackage{verbatim}
\usepackage{cancel}
\usepackage{chemarrow}
\usepackage[linktocpage=false,colorlinks = true,
            linkcolor = darkviolet,
            urlcolor  = blue,
            citecolor = blue,
            anchorcolor = magenta]{hyperref}

\definecolor{cream}{rgb}{.97, .95, .88}
\definecolor{darkcream}{rgb}{1., .88, .5}
\definecolor{lightpink}{rgb}{0.98, 0.88, 0.87}
\definecolor{lightwhite}{rgb}{1., 0.98, 0.95}
\definecolor{lightsalmon}{rgb}{1., 0.95, 0.90}
\definecolor{lightviolet}{rgb}{0.9, 0.8, 0.9}
\definecolor{lightgray}{rgb}{.96, .96, .96}  
\definecolor{lgray}{rgb}{.75, .75, .75}
\definecolor{LemonChiffon}{rgb}{0.95, 1., 0.7}
\definecolor{lightolivegreen}{rgb}{0.84, 0.89, 0.25}
\definecolor{lightgreen}{rgb}{.664, 1., .52}
\definecolor{llgreen}{rgb}{.900, .983, .960}
\definecolor{tristle}{rgb}{0.87, 0.67, 0.77} 
\definecolor{pink}{rgb}{0.95, 0.45, 0.75}
\definecolor{magenta}{rgb}{1., 0, 1.}
\definecolor{violet}{rgb}{0.9, 0.20, 0.85}
\definecolor{darkolivegreen}{rgb}{0.55, 0.65, 0.35}
\definecolor{maroon}{rgb}{0.7, 0.26, 0.56}
\definecolor{lightmaroon}{rgb}{0.85, 0.38, 0.58}
\definecolor{darkmaroon}{rgb}{0.604, 0.169, 0.451}
\definecolor{ddarkmaroon}{rgb}{0.2, 0.03125, 0.150}
\definecolor{mediumorchid}{rgb}{0.8, 0.33, 0.83}
\definecolor{mediumorchidd}{rgb}{1., 0.33, 0.63}
\definecolor{darkgreen}{rgb}{0.1, 0.6, 0.13}
\definecolor{lightyellow}{rgb}{1., 1., 0.82}
\definecolor{turquoise}{rgb}{0.042, 0.586, 0.512}
\definecolor{turquoisel}{rgb}{0.66, 0.94, 0.83}
\definecolor{darkturquoise}{rgb}{0.21, 0.55, 0.50}
\definecolor{coral}{rgb}{1., 0.6, 0.21}
\definecolor{lightorange}{rgb}{1., 0.88, 0.75}
\definecolor{orangered}{rgb}{1., 0.5, 0.}
\definecolor{orange}{rgb}{1., 0.65, 0.1}
\definecolor{orangel}{rgb}{1., .85, .3}
\definecolor{darkorange}{rgb}{0.875, 0.4, 0.204}
\definecolor{ddarkorange}{rgb}{.675, .218, .05}
\definecolor{bluesky}{rgb}{0.48, 0.53, 1.}
\definecolor{gold}{rgb}{1., 0.85, 0.25}
\definecolor{goldd}{rgb}{0.95, 0.75, 0.05}
\definecolor{darkviolet}{rgb}{0.54, 0.04, 0.84}
\definecolor{ddarkviolet}{rgb}{.382, .063, .657}
\definecolor{lightblue}{rgb}{0.30, 0.86, 0.89}
\definecolor{LightBlue}{rgb}{0.68, 0.85, 0.9}
\definecolor{lblue}{rgb}{0.78, 0.90, 0.95}
\definecolor{darkblue}{rgb}{.105, .308, .707}
\definecolor{lightmaroon}{rgb}{0.85, 0.38, 0.58}
\definecolor{darkmaroon}{rgb}{0.604, 0.169, 0.451}
\definecolor{darkpink}{rgb}{0.879, 0.020, 0.766}
\definecolor{ddarkpink}{rgb}{0.738, 0.195, 0.406}
\definecolor{grey}{rgb}{0.717, 0.717, 0.717}
\definecolor{lightgrey}{rgb}{0.800, 0.800, 0.800}
\definecolor{brown}{rgb}{0.740, 0.323, 0.182}
\definecolor{redbrown}{rgb}{.575, .158, .05}
\definecolor{darkbrown}{rgb}{0.34, 0.25, 0.05}
\definecolor{orangebrown}{rgb}{0.433, 0.262, 0.06}
\definecolor{pinkl}{rgb}{1., 0.788, 0.918}
\definecolor{salmon}{rgb}{1., 0.66, 0.5}
\definecolor{lightbrown}{rgb}{0.703, 0.508, 0.121}

\def\Journal#1#2#3#4{{#1} {\bf #2}, (#3) #4}

\def\Name#1#2 {{#2, }{#1}}

\def\AA{\em A.\& A.}

\def\APH{\em Annals Phys.}

\def\ATM{\em Adv. Theor. Math. Phys}

\def\CQG{\em Class. Quant.~Grav.}

\def\FOP{\em Found. Phys.}

\def\GCS{\em Grav. Cosmol. Suppl.}
\def\GRG{\em Gen. Rel. Grav}

\def\IMA{{\em Int. J. Mod. Phys.} \emph{A}}
\def\IMD{{\em Int. J. Mod. Phys.} \emph{D}}

\def\ITP{\em Int. J. Theor. Phys.}
\def\JCA{\em J. Cosmol. Astrop. Phys.}

\def\JHE{\em J. High Energy Phys.}
\def\JMP{\em J. Math. Phys.}
\def\JPA{\em J.~Phys.~A}
\def\JPC{\em J. Phys. Conf. Ser.}

\def\JPU{\em J. Phys. (USSR)}

\def\MPL{{\em Mod. Phys. Lett.} \emph{A}}

\def\NAT{\em Nature}

\def\NPB{{\em Nucl.~Phys.}~\emph{B}}

\def\PLA{{\em Phys. Lett.}~\emph{A}}
\def\PLB{{\em Phys. Lett.}~\emph{B}}

\def\PRD{{\em Phys.~Rev.}~\emph{D}}
\def\PRL{\em Phys. Rev. Lett.}
\def\PRV{\em Phys.~Rev.}
\def\PRE{\em Phys.~Rep.}

\def\PRLA{{\em Proc. Roy. Soc. Lond.}~\emph{A}}

\def\SCI{\em Science}

\def\be{\begin{equation}}
\def\ee{\end{equation}}
\def\bea{\begin{eqnarray}}
\def\eea{\end{eqnarray}}
\def\bes{\begin{equation*}}
\def\ees{\end{equation*}}
\def\beas{\begin{eqnarray*}}
\def\eeas{\end{eqnarray*}}

\def\tr{\text{tr}}

\def\cf{\mathtt f}
\def\bm{\mathcal B}

\def\hm{\mathcal H}

\def\hH{\hat{H}}
\def\hJ{\hat{J}}
\def\hL{\hat{L}}
\def\hO{\hat{O}}

\def\hT{\hat{T}}

\begin{document}
\begin{center}
{\Large \bf Making a Quantum Universe: Symmetry and Gravity}\\
\end{center}

\begin{center}
Houri Ziaeepour$^{a,b}$\footnote{Email: {\tt houriziaeepour@gmail.com}} \\
\end{center}
$^a$Institut UTINAM, CNRS UMR 6213, Observatoire de Besan\c{c}on, Universit\'e de Franche Compt\'e, 
41 bis ave. de l'Observatoire, BP 1615, 25010 Besan\c{c}on, France \\
$^b$Mullard Space Science Laboratory, University College London, Holmbury St. Mary, 
GU5 6NT, Dorking, UK


\begin{abstract}
So far, none of attempts to quantize gravity has led to a satisfactory model that not only describe 
gravity in the realm of a quantum world, but also its relation to elementary particles and other 
fundamental forces. Here, we outline the preliminary results for a model of quantum universe, in which 
gravity is fundamentally and by construction quantic. The model is based on three well motivated 
assumptions with compelling observational and theoretical evidence: quantum mechanics is valid at all 
scales; quantum systems are described by their symmetries; universe has infinite independent degrees of 
freedom. The last assumption means that the Hilbert space of the Universe has 
$SU(N\rightarrow \infty) \cong \text{area preserving Diff.} (S_2)$ symmetry, which is 
parameterized by two angular variables. We show that, in the absence of a background spacetime, this 
Universe is trivial and static. Nonetheless, quantum fluctuations break the symmetry and divide the 
Universe to subsystems. When a subsystem is singled out as reference---observer---and another as clock, 
two more continuous parameters arise, which can be interpreted as distance and time. We identify the 
classical spacetime with parameter space of the Hilbert space of the Universe. Therefore, its quantization 
is meaningless. In this view, the Einstein equation presents the projection of quantum dynamics in the 
Hilbert space into its parameter space. Finite dimensional symmetries of elementary particles emerge as 
a consequence of symmetry breaking when the Universe is divided to subsystems/particles, without having 
any implication for the infinite dimensional symmetry and its associated interaction - perceived as gravity. 
This explains why gravity is a universal force.
\end{abstract}
\pagebreak
\tableofcontents

\section{Introduction and Summary of~Results}
More than a century after the discovery of general relativity and description of gravitational force as the 
modification of spacetime geometry by matter and energy, we still lack a convincing model for explaining these 
processes in the framework of quantum mechanics. Appendix~\ref{app:qgrsumm} briefly reviews the history of 
efforts for finding a consistent Quantum Gravity (QGR) model. Despite tremendous effort of generations of 
scientists, none of proposed models presently seem fully~satisfactory. 

Quantization of gravity is inevitable. Examples of inconsistencies in a universe where matter is ruled by 
quantum mechanics but gravity is classical are well known~\cite{grinconsist,houriqgr}. In~addition, 
in~\cite{houriqgr}, it is argued that there must be an inherent relation between gravity and quantum 
mechanics. Otherwise, the~universality of Planck constant $\hbar$ as quantization scale cannot be 
explained.\footnote{Nonetheless, Ref.~\cite{planckcontex} advocates a context dependent Planck constant.}
Aside from these arguments, the~fact that there is no fundamental mass/energy scale in quantum mechanics 
means that it has to have a close relation with gravity that provides a dimensionful fundamental constant, 
namely the Newton gravitational constant $G_N$ or equivalently the Planck mass 
$M_P \equiv \sqrt{\hbar c / G_N}$, where $c$ is the speed of light in vacuum 
(or equivalently Planck length scale $L_P \equiv \hbar /cM_P$). We should remind 
that a dimensionful scale does not arise in conformal or scale independent models. Indeed, conformal 
symmetry is broken by gravity, which provides the only fundamental dimensionful constant to play the 
role of a ruler and make distance and mass measurements meaningful.\footnote{In addition to $M_P$, we 
need two other fundamental constants to describe physics and cosmology: the Planck constant $\hbar$ and 
maximum speed of information transfer that experiments show to be the speed of light in classical vacuum. 
We remind that triplet constants $(\hbar, c, M_P)$ are arbitrary and can take any nonzero positive value. 
The~selection of their values amounts to the definition of a system of units for measuring other physical 
quantities. In~QFT literature usually $\hbar = 1$ and $c = 1$ are used. In~this system of units---called high 
energy physics units~\cite{earlyuniv}---$\hbar$ and $c$ are dimensionless.}

In what concerns the subject of this volume, namely representations of inhomogeneous Lorentz symmetry (called 
also Poincar\'e group), they were under special interest since decades ago, hoping that they help formulate 
gravity as a renormalizable quantum field. The~similarity of the compact group of local Lorentz transformations to 
Yang--Mills gauge symmetry has encouraged quantum gravity models that are based on the first order formulation of 
general relativity. These models use vierbein formalism and extension of gauge group of elementary particles 
to accommodate Poincar\'e group~\cite{qgrgauge,qgrgauge0}. However, Coleman--Mandula theorem~\cite{ssymmtheorem} 
on $S$-matrix symmetries---local transformations of interacting fields that asymptotically approach 
Poincar\'e symmetry at infinity---invalidates any model in which Poincar\'e and internal symmetries are not 
factorized. According to this theorem total symmetry of a grand unification model, including gravity, must be 
a tensor product of spacetime and internal symmetries. Otherwise, the~model must be 
supersymmetric~\cite{ssymmsusy} or VEV of the gauge field should not be flat~\cite{grgaugemix}. However, 
we know that even if supersymmetry is present at $M_P$ scale, it is broken at low energies. Moreover, any 
violation of Coleman--Mandula theorem and Lorentz symmetry at high energies can be convoyed to 
low energies~\cite{lqglorentz} and violate e.g.,~equivalence principle and other 
tested predictions of general relativity~\cite{grtest,grestgrb090510a}. For~these reasons, modern approaches 
to the unification of gravity as a gauge field with other interactions consider the two sectors as separate 
gauge field models. In~addition, in~these models gravity sector usually has topological action to make 
the formulation independent of the geometry of underlying spacetime, see 
e.g.,~\cite{qgrgaugesep,qgrgaugesep0,qgrgaugesep1}. However, like other quantum gravity candidates these models
suffer from various issues. The~separation of internal and gravitational gauge sectors means that these models 
are not properly speaking a grand unification. Moreover, similar to other approaches to QGR, these models do 
not clarify the nature of spacetime, its dimensionality, and~relation between gravity and internal~symmetries.

In addition to consistency with general relativity, cosmology, and~particle physics, a~quantum model unifying 
gravity with other forces is expected to solve well known problems that are related to gravity and spacetime, 
such as: physical origin of the arrow of time; apparent information loss in black holes; and, UV and IR 
singularities in Quantum Field Theory (QFT) and general gravity.\footnote{Some quantum gravity models such 
as loop quantum gravity emphasize the quantization of gravity alone. However, giving the fact that gravity 
is a universal force and interacts with matter and other forces, its quantization necessarily has impact on 
them. Therefore, any quantum gravity only model would be, at~best, incomplete.} There are also other issues 
that a priori should be addressed by a QGR model, but~are less discussed in the~literature:
\begin{enumerate}
\item {\bf Should spacetime be considered as a physical entity similar to quantum fields associated to particles, or~rather it presents a configuration space ?} \label{spacephys}
\begin {description}
\item General relativity changed spacetime from a rigid entity to a deformable media. However, it does not specify 
whether spacetime is a physical reality or a property of matter, which~ultimately determines its geometry and 
topology. We remind that in the framework of QFT vacuum is not the empty space of classical physics, 
see e.g.,~\cite{qmcurve,qmcurve0}. In~particular, in~the presence of gravity the naive definition of quantum vacuum 
is frame dependent. A~frame-independent definition exists~\cite{hourivacuum} and it is very far from classical 
concept of an empty space. Explicitly or implicitly, some of models reviewed in Appendix~\ref{app:qgrsumm} 
address this question. 
\end {description}
\item {\bf Is there any relation between matter and spacetime ?} \label{matterspace} 
\begin {description}
\item In general relativity matter modifies the geometry of spacetime, but~the two entities are considered as 
separate and stand alone. In~string theory spacetime and matter fields---compactified internal space---are 
considered and treated together, and~spacetime has a physical reality that is similar to matter. By~contrast, 
many other QGR candidates only concentrate their effort on the quantization of spacetime and gravitational 
interaction. Matter is usually added as an external ingredient and it does not intertwine in the construction 
of quantum gravity and spacetime.
\end {description}
\item {\bf Why do we perceive the Universe as a three-dimensional (3D) space (plus time) ?} \label{spacedim}
\begin {description}
\item None of extensively studied quantum gravity models discussed in Appendix~\ref{app:qgrsumm} answer 
this question, despite the fact that it is the origin of many troubles for them. For~instance, the~enormous 
number of possible models in string theory is due to the inevitable compactification of extra-dimensions to 
reduce the dimension of space to the observed 3 + 1. In~background independent models, the~dimension of space 
is a fundamental assumption and essential for many technical aspects of their construction. In~particular, 
the definition of Ashtekar variables~\cite{ashtekarvar} for $SU(2) \cong SO(3)$ symmetry and its relation 
with spin foam description of loop quantum gravity~\cite{lqgfoam} are based on the assumption of a 3D real 
space. On~the other hand, according to holography principle, the~maximum amount of information that is containable 
in a quantum system is proportional to its area rather than volume. If~the information is projected and 
available on the boundary, it is puzzling why we should perceive the volume.
\end {description}
\end {enumerate}

In a previous work~\cite{houriqmsymm}, we advocated the foundational role of symmetries in quantum mechanics 
and reformulated its axioms accordingly, see Appendix~\ref{app:qmaxioms} for a summary. Of~course, the~
crucial role of symmetries in quantum systems is well known. However, axioms of quantum mechanics \`a la Dirac 
and von Neumann consider an abstract Hilbert space and do not specify its relation with symmetries of 
quantum systems. In~addition to symmetries of their classical Lagrangian, Hilbert space of quantum systems 
represents $SU(N)$ group, called state symmetry, see Appendix~\ref{app:cohersymm} for more details. 
Transformation of states by this group modifies their coherence, and~recently quantification of this 
property and its usefulness as a resource has become a subject of interest in quantum information theory 
literature~\cite{qminfocohere,qmspeedlimit}. 

Inspired by these developments, in~this work we study a standalone quantum system, which is considered to be the 
Universe.

\subsection{Summary of the Model and Results} \label{sec:summary}
The model assumes infinite number of independent and simultaneously commuting observables in the Universe, 
but no background spacetime. Hilbert space $\hm_U$ of such system represents 
$SU(N \rightarrow \infty)$ symmetry. However, in absence of a background spacetime, its dynamics is 
trivial and its Lagrangian is defined on the group manifold of $SU(\infty)$ symmetry. Therefore, states are 
pure gauge. The~vector space of gauge transformations, corresponding to linear transformations of the 
Hilbert space, is $\bm[\hm_U] \cong SU(\infty)$. On~the other hand, quantum fluctuations break the state symmetry and factorize the Hilbert space to blocks of tensor product of subspaces according to criteria studied 
in~\cite{sysdiv,targetspacebreak}. For~each subsystem, the~rest of the Universe plays the role of a background parameterized by three continuous quantities that can be identified with the classical space. 
Moreover, division of the Universe to subsystems leads to emergence of time and its arrow \`a la 
Page \& Wootters~\cite{qmtimepage} or similar methods~\cite{qmtimedef}. We show that the 3 + 1 dimensional 
parameter space is, in~general, curved and invariant under inhomogeneous Lorentz transformations and its 
curvature is determined by quantum states of the subsystems. We also comment on the signature of parameter 
space metric. Based on these observations, we interpret $SU(\infty)$ sector of the model as Quantum Gravity. The~finite rank factorized symmetries become local gauge fields acting on a Hilbert 
space that presents matter~fields.

These results demonstrate the importance of the division of Universe to subsystems and the distinction of 
observer and clock from the rest. Nonetheless, in~contrast to the Copenhagen interpretation of quantum 
mechanics, the~absence of observer does not make the model meaningless, but~trivial and static. This model 
answers some of issues raised in questions \ref{spacephys}--\ref{spacedim} raised earlier. In~particular, 
it~clarifies the nature of spacetime and its dimensionality, and provides an explanation for the universality 
of gravitational force.

A crucial proposal of the model is that what we perceive as classical spacetime is the configuration 
(parameter) space of its content. In~other words, rather than saying particles/objects (such as strings) 
live in a 3 + 1 dimensional space, according to this model we can say that an ensemble of abstract objects with 
$SU (\infty) \times G$ symmetry look like a 3 + 1 dimensional infinite curved spacetime with gravity, 
where~subsystems are fields that represent group $G$ as a local gauge symmetry. Thus,~we~can completely neglect the 
geometric interpretation and just consider the Universe as an infinite tensor product. This~aspect of the 
model is similar to the approach of~\cite{qgrentangle1}. However, their model is somehow inverse of that 
studied here. They use tensor product and quantum entanglement to make a symplectic geometry that becomes a 
continuous curved spacetime when the number of tensor product factors approaches to infinity. The~drawback is 
that symplectic geometries defined by graphs can be embedded in any space of dimension $D \geqslant 2$. 
Consequently, they~cannot explain 
the dimension of the~spacetime.

Axioms and structure of the model is discussed in Section~\ref{sec:infuniverse}. Lagrangian of the system 
before its division is described in Section~\ref{sec:lagrangian}. Properties of the model after symmetry breaking 
and division of the Universe are studied in Section~\ref{sec:division}. Section~\ref{sec:compmodel} presents 
a brief comparison of this model with string and loop quantum gravity. Section~\ref{sec:outline} presents 
outlines and prospective for future investigations. Accompanying appendices contain technical details 
and review of previous results. Appendix \ref{app:qgrsumm} gives a short recount of the history of quantum 
gravity models. Appendix \ref{app:qmaxioms} summaries the axioms of quantum mechanics in symmetry language. 
State space and its associated symmetry is reviewed in Appendix \ref{app:cohersymm}. Properties of 
$SU(\infty)$ and its representations are summarized in Appendix \ref{app:harmonicdecomp} and its Cartan 
decomposition in Appendix \ref{app:cartandecomp}.

\section {An Infinite Quantum Universe} \label{sec:infuniverse}
Our departure point for constructing a quantum universe consists of three well motivated assumptions 
with compelling observational and theoretical evidence: 
\setcounter{enumi}{0}
\renewcommand{\theenumi}{\Roman{enumi}}
\begin{enumerate}
\item Quantum mechanics is valid at all scales and applies to every entity, including the Universe as 
a whole; \label{qmunivassum}
\item Any quantum system is described by its symmetries and its Hilbert space represents them; \label {symmassum}
\item The Universe has an infinite number of independent degrees of freedom. \label{infassum} 
\end{enumerate}

The last assumption means that the Hilbert space of the Universe $\hm_U$ is infinite dimensional and 
represents the group $SU(\infty)$. There is sufficient evidence in favour of such an assumption. For~instance, 
the~thermal distribution of photons at IR limit contains an infinite number of quanta with energies approaching 
zero and there is no minimum energy limit. For~this reason, vacuum can be considered to be a superposition of 
multi-particle states of any type---not just photons---without any limit on their number~\cite{hourivacuum}. 
In general relativity, there is no upper limit for gravitons wavelength and thereby their number. 
Of course, one may argue that a lower limit on energy or spacetime volume may exist. Nonetheless, for~any 
practical application the number of subsystems/quanta in the Universe can be considered to approach infinity. 
Indeed, even in quantum gravity models that assume a symplectic structure for spacetime, such as spin 
foam/loop quantum gravity and causal sets, there is not a fixed lattice of spacetime and the number of spacetime 
states is effectively~infinite.

The algebra that is associated to the $SU(\infty)$ coherence (state) symmetry of the above model is defined 
as\footnote{In this work, all vector spaces and algebras are defined on complex number field 
$\mathbb{C}$, unless~explicitly mentioned otherwise.}:
\be
[\hL_a,\hL_b] = \frac{\hbar}{cM_P} f_{ab}^c \hL_c = L_P f_{ab}^c \hL_c \label {statealgebra}
\ee
where operators $\hL_\alpha \in \bm[\hm_U]$ are generators of algebra $su(\infty)$ and $f_{ab}^c$ are its 
structure coefficients. They~are normalized such that the r.h.s. of (\ref{statealgebra}) explicitly 
depends on the Planck constant $\hbar$. If~$\hbar \rightarrow 0$, the~r.h.s. becomes null, and the algebra 
becomes abelian and homomorphic to $\bigotimes^{N \rightarrow \infty} U(1)$ --- in~agreement with the symmetry of 
configuration space of classical systems, explained in Appendix \ref{app:cohersymm}. The~same happens 
if $M_P \rightarrow \infty$, that is when Planck mass scale is much larger than scale of interest. 
In both cases $L_P \rightarrow 0$. Assuming that $SU(\infty)$ symmetry of the Universe can be associated 
to gravitational interaction---we will provide more evidence in favour of this claim later---the above 
limits mean that, in~both cases, gravity becomes negligible.\footnote{Although in (\ref{statealgebra}) we show 
the dimensional scale $\hbar/M_P$ in the definition of operators and their algebra, for~the sake of 
convenience in the rest of this work, we include it in the operators, except~when its explicit presentation 
is necessary for the~discussion.}

It is well known that 
$\bm[\hm_U] \cong SU(\infty) \cong \text{area preserving Diff}(S_2)$~\cite{suninfhoppthesis,suninfym}, where 
$S_2$ is 2D sphere. In~fact, $SU(\infty)$ is homomorphic to area preserving diffeomorphism of any 
two-dimensional (2D) Riemann surface~\cite{suninftorus,suninfrep,suninfrep0}. Therefore, here $S_2$ can be 
any 2D surface, rather than just sphere. This~theorem can be heuristically understood as the following: any 
compact 2D Riemann surface can be obtained from sphere by removing a measure zero set of pairs of points and 
sticking the rest of the surface pair-by-pair together. Although~surfaces with different genus are topologically 
different, they~are homomorphic. This property may be important in the presence of subsystems with singularity, 
such as black holes, in~which part of the parameter space is inaccessible. From~now on, we call a 2D surface 
that its diffeomorphism represents $SU(\infty)$ a~diffeo-surface. 

Homomorphism between $SU(\infty)$ and $\text{Diff}(S_2)$ makes it possible to expand $\hL_a$'s with respect 
to spherical harmonic functions, depending on angular coordinates $(\theta, \phi)$ on a sphere. Moreover, 
owing to the Cartan decomposition, $SU(\infty)$ generators can be described as a tensor product of Pauli 
matrices~\cite{suninfhoppthesis,cartandecomp}. In~this case, indices in (\ref{statealgebra}) consist of a 
pair $(l,m)~|~l = 0, \cdots, \infty; -l \leqslant m \leqslant +l$. Appendix~\ref{app:cartandecomp} reviews 
decomposition and indexing of $SU(\infty)$ generators. We continue to use single letters for the indices of 
generators when there is no need for their explicit~description. 

The algebra (\ref{statealgebra}) is not enough to make the system quantic and as usual $\hat{L}_a$'s must 
respect Heisenberg commutation relations:
\be
[\hL_a,\hJ_b] = -i \delta_{ab} \hbar. \label{lquantize}
\ee
where $\hJ_a \in \bm[\hm_U^*]$ is the dual of $\hL_a$ and $\hm_U^*$ is the dual Hilbert space of the Universe. 
As there is a one-to-one correspondence between $\hL$'s and $\hJ$'s, they satisfy the same algebra, 
represent the same symmetry group, namely $SU(\infty)$, and~they have their own expansion to spherical harmonics. 
Owing to $SU(\infty) \cong \text{Diff}(S_2)$, vectors of the Hilbert space are differentiable complex 
functions of angular coordinates $(\theta, \phi)$. Thus, spherical harmonic functions constitute an 
orthogonal basis for $\hm_U$. The~Cartan subalgebra of $\bm [\hm_U] \cong SU(\infty)$ is also infinite 
dimensional.

The quantum Universe defined here is static, because~there is no background space or time in the model. 
Nonetheless, in~Section \ref {sec:division}, we show that continuous degrees of freedom similar to space and time naturally arise when the Universe is divided to subsystems. The~short argument goes as the~following: 

We assume that eigen states of the Hilbert space of the Universe are not abstract objects and physically exist.
This assumption is supported by the fact that in Standard Model (SM) of particle physics states that constitute a 
basis for its Hilbert space and for its space of linear transformations are indeed observed particles (fields). 
Consequently, taking into account the assumption that $\hm_U$ is infinite dimensional, we conclude that the 
Universe must consist of infinite number of particles/subsystems. Although~subsystems may have some  
common properties, which make them indistinguishable from each others, there are many other distinguishable 
aspects, which discriminate them from each others. This statement is in agreement with the corollary presented in 
Appendix \ref{app:qmaxioms} regarding the divisibility of a quantum Universe, derived from axioms of quantum 
mechanics. Thus, this conclusion is independent of details of the~model. 

Notice that, without~the assumption about physical existence of eigen states, an~infinite dimensional Hilbert 
space does not necessarily mean Universe must be infinitely divisible. Hilbert~space of many quantum 
systems have an infinite number of states. However, they do not necessarily occur in each instance (copy) of the 
system. The~case of a Universe is different, because, by~definition, there is only one copy of it. Therefore, 
every eigen state of a complete basis of its Hilbert space must physically exist. Otherwise, it can be 
completely~discarded.

In the next sections, we make this argument more rigorous and explain how it can lead to a 3 + 1 dimensional 
spacetime and internal gauge symmetry of elementary particles. We begin with constructing a Lagrangian 
for this static model and show that it is~trivial.

\section {Lagrangian of the Universe} \label{sec:lagrangian}
Although the infinite dimensional Universe described in the previous section is static, it has to satisfy 
constraints imposed by symmetries associated to it. They are analogous to constraints imposed 
on systems in thermodynamic equilibrium. Although~there is no time variation in such systems, a~priori 
small perturbations occur, for~instance, by~absorption and emission of energy. They must be in balance with 
each others, otherwise the system would lose its equilibrium. Therefore, it is useful to define a Lagrangian 
that quantifies these constraints. In~the case of present model, the~Lagrangian should quantify 
$SU(\infty)$ symmetry and its representation by $\hm_U$.

Lagrangian of a system must be invariant under transformations of fields by application of members of its  
symmetry group. As~there is no background spacetime in this model, the~most appealing candidate is a 
Lagrangian similar to Yang--Mills, but~without a kinetic term. In~such a situation, the~only available 
quantities are invariants of the symmetry group:
\be
\mathcal {L}_U = \int d^2\Omega \sqrt{|g^{(2)}|} \biggl [\frac{1}{2} \sum_{a,~b} L^*_a(\theta,\phi) 
L_b (\theta, \phi) \tr (\hL_a \hL_b ) + \frac {1}{2} \sum_a L_a \tr (\hL_a \rho (\theta, \phi)) \biggr ], 
\quad \quad d^2\Omega \equiv d(\cos \theta) d\phi \label{twodlagrang}
\ee
where $g$ is the determinant of 2D metric of the diffeo-surface. If~we use description (\ref{lharminicexp}) 
for $\hL$ operators, $a=b=1$. If~we use (\ref{xcommute}) expression,  
$a,b = (l,m),~l=0, \cdots , \infty; -m \leqslant l \leqslant +m$. The~latter case explicitly demonstrates 
the Cartan decomposition of $SU(\infty)$ to $SU(2)$ factors, as~described in Appendix~\ref{app:cartandecomp}. 
Notice that $(\theta,\phi)$ are internal variables~\cite{suninfym}, reflecting the fact that vectors 
of the Hilbert space representing $SU(\infty)$ are functions on a 2D Riemann surface. For~the 
same reason, in~contrast to usual Lagrangians in QFT, there is no term containing derivatives with respect to 
these parameters in $\mathcal {L}_U$. If~we use differential representation of $\hL_{lm}$ defined in 
(\ref{lharminicexp}) and apply it to amplitudes $L_{lm} (\theta,\phi)$, the~first term in the Lagrangian will 
depend on the partial derivatives of amplitudes, just like in the QFT. However, it is straightforward to see that 
derivatives with respect to $\cos \theta$ and $\phi$ will have different amplitudes and, thereby, the~
kinetic term will be unconventional and non-covariant, unless~we consider amplitudes 
$L_{lm} (\theta,\phi)$ as functions of the metric of a deformed sphere. This is the explicit demonstration of 
$SU(\infty) \cong \text{Diff}(S_2)$ invariance of this~Lagrangian.

Generators $\hT_a, \hT_b \in SU(N), \forall N$ can be normalized, such that 
$\tr (T_a T_b) ~ \propto ~ \delta_{ab}$, see e.g.,~\cite{suninfhoppthesis}. In~analogy with field strength in 
Yang--Mills theories, the~function $L_a (\theta, \phi)$ can be interpreted as the amplitude of the 
contribution of operator $\hL_a$ in the dynamics of the Universe. Due to global $U(1)$ symmetry of operators 
applied to a quantum state, $L_a$'s are, in~general, complex. On~the other hand, when considering the Cartan 
decomposition of $SU(\infty)$ to tensor product of $SU(2)$ factors and the fact that 
$\sigma^\dagger = (\sigma*)^t = \sigma$, we conclude that $\hL_a^\dagger = \hL_a$, Similar to QFT, one can use 
$\mathcal {L}_U$ to define a path integral. In~the absence of time, the~path integral presents the excursion 
of states in the Hilbert space by successive application of $\hL_a$ operators. Nonetheless, owing to 
$SU(\infty)$ symmetry, variation of states is equivalent to gauge transformation and~non-measurable. 

The analogy of $\mathcal {L}_U$ with Yang--Mills theory has interesting consequences. For~instance, 
differential representation of $\hL_{lm}$ defined in (\ref{lharminicexp}) can be written as 
$\hL_{lm} = \sqrt{|g^{(2)}|} \epsilon^{\mu\nu} (\partial_\mu Y_{lm}) \partial_\nu$. In~classical limit, one can 
consider that $\hL_{lm}$ acts on the field amplitude $L_{lm}$ and the first term in the integrand of Lagrangian 
$\mathcal {L}_U$ can be arranged, such that it becomes proportional to Ricci scalar $R^{(2)}$. As~the 
geometry of 2D diffeo-surface is arbitrary, for~each set of $L_{lm}$ the metric $g_{\mu\nu}$ can be chosen, such 
that $L_{lm}$ dependent part of the integrand becomes proportional to Ricci scalar for that metric. 
Thus, in~classical limit the first term is topological.\footnote{We remind that 
$\int_{\mathcal{M}} d^2\Omega \sqrt{|g^{(2)}|} R^{(2)} = 4\pi \chi (\mathcal{M})$, where $\chi$ is the Euler 
characteristic of the compact Riemann 2D surface $\mathcal{M}$. Moreover, Ricci scalar alone does not 
determine Riemann curvature tensor $R_{\mu\nu}$ and only provides one constraint for three independent 
components of the metric tensor.} 
We could arrive to this conclusion inversely. Because~$SU(\infty) \cong \text{Diff}(S_2)$, in~the classical 
limit the Lagrangian should be the same as Einstein 
gravity in a static 2D curve space. Thus, the~first term in (\ref {twodlagrang}) can be replaced by 
$\int d^2\Omega \sqrt{|g^{(2)}|} R^{(2)}$. Then, the~definition of $\hL_{lm}$ operators in~(\ref{lharminicexp}) 
and amplitudes $L_{lm}$ can be used to write $R^{(2)}$ with respect to $\hL_{lm}$ and relate metric and 
connection of the 2D surface to amplitudes $\L_{lm}$. We leave a detailed demonstration of these relations 
to a future work. The~relation between gauge field term in $\mathcal {L}_U $ and Riemann curvature in 
classical limit is crucial for interpretation of this term as gravity when the Universe is divided to 
subsystems.

Notice that, in~both representations of $SU(\infty)$, namely Cartan decomposition to tensor product of $SU(2)$ 
factors and diffeomorphism of 2D surfaces, angular coordinates $\theta$ and $\phi$ play the role of 
parameters that identify/index the members of the symmetry group. Consequently, their quantization is 
meaningless. This is consistent with interpretation of Einstein equation as an equation of state~\cite{greos}. 
Presuming the physical reality of Hilbert space and operators applied to it, as~discussed in 
the previous section and in Appendix \ref{app:cohersymm}, we can interpret $L_{lm}$ as intensity of force 
mediator particles related to the symmetry represented by operators $\hL_{lm}$, and~$\rho$ in the 
second term of the Lagrangian $\mathcal {L}_U$ as density matrix of~matter. 

Although $\mathcal {L}_U$ is static, we can apply a variational principle with respect to amplitudes to obtain 
field equations and find equilibrium values of $L_{lm}$ and $\rho$. However, it is easily seen that 
solutions of these equations are trivial. At~equilibrium $L_{lm} \rightarrow 0$ and $\rho_{lm} \rightarrow 0$, 
see Appendix \ref{app:dyneq} for the details. Because~$SU(\infty)^n \cong SU(\infty)~ \forall ~n$, this 
solution has properties of a frame independent vacuum of a many-particle Universe defined by using 
coherent states~\cite{hourivacuum}. Their similarity implicitly implies that the Universe is divisible and 
consists of infinite number of particles/subsystems interacting through mediator particles of 
$SU(\infty)$ force, which is the action of $\hL_{lm}$. We investigate this conclusion in more details in 
the next~section.

\section {Division to Subsystems} \label{sec:division}
There are many ways to see that the quantum vacuum (equilibrium) solution of a Universe with $\mathcal{L}_U$ 
Lagrangian (\ref{twodlagrang}) is not stable. Of~course there are quantum fluctuations. They are nothing 
else than random application of $\hL_{lm}$ operators, in~other words random scattering of force mediator 
quanta by matter. They project the Hilbert space to itself. However, owing to $SU(\infty)$ symmetry of 
Lagrangian, states are globally equivalent and the Universe maintain its equilibrium. Nonetheless, locally 
states are different and they do not respond to $\hL_{lm}$ in the same manner. Here, locality means restriction 
of Lagrangian and projections to a subspace of the Hilbert space~\cite{targetspacebreak}. As~state space is 
homomorphic to the space of smooth functions on the sphere $f (\theta,\phi)$, the~restriction of transformations 
to a subspace is equivalent to a local deformation of the diffeo-surface. Moreover, the~difference between 
structure coefficients of $SU(\infty)$ can be used to define a locality or closeness among operators that 
belong to $\bm[\hm_U]$. These observations are additional evidence to the argument given at the end of 
Section~\ref{sec:lagrangian} in favour of the divisibility of the quantum Universe introduced in 
Section~\ref{sec:infuniverse} to multi-particle/subsystems.

A quantum system that is divisible to separate and distinguishable subsystems \footnote{In statistical quantum 
or classical mechanics distinguishability of particles usually means being able to say, for~instance, whether 
it was particle 1 or particle 2 which was observed. Here by distinguishability we mean whether a 
particle/subsystem can be experimentally detected, i.e.,~through application of $\hL_{lm}$ to a subspace of 
parameter space and identified in isolation from other subsystems or the rest of the Universe.} must 
fulfill 3 conditions~\cite{sysdiv}:
\begin{enumerate}
\item[-] There must exist sets of operators $\{A_i\} \subset \bm [\hm]$ such that 
$\forall \bar{a} \in \{A_i\}$ and~$\forall \bar{b} \in \{A_j\},$ and~$i \neq j,~ [\bar{a}, \bar{b}] = 0$; 
\item[-] Operators in each set $\{A_i\}$ must be local\footnote{This condition 
is defined for quantum systems in a background spacetime. In~the present model there is not such a 
background. Nonetheless, as~explained earlier, locality on the diffeo-surface can be projected to 
$\bm [\hm_U]$.};
\item[-] $\{A_i\}$'s must be complementarity, which is $\otimes_i \{A_i\} \cong \text{End} (\bm [\hm])$. 
\end{enumerate}

The most trivial way of fulfilling these conditions is a reducible representation of symmetries 
by $\bm [\hm]$. In~the case of $\bm [\hm_U] \cong SU(\infty)$, as:
\be
SU(\infty)^n \cong SU(\infty) ~\forall ~n \label{multysuinf}
\ee 
the above condition can be easily realized. Moreover, instabilities, quantum correlations, and~entanglement 
may create local symmetries among groups of states and/or operators. There~are many examples of such 
grouping and induced symmetries in many-body systems, see e.g.,~\cite{anyonrev} for a review. A~hallmark of 
induced symmetry by quantum correlations is the formation of anyon quasi-particles having non-abelian symmetry 
in the fractional quantum Hall effect~\cite{fracqmhall}. On~the other hand, there is only one state in the 
infinite dimensional Hilbert space, in~which all pointer states have the same probability, namely the maximally 
coherent state defined in (\ref{maxcoherestate}). Even if a many-body system begins in such a maximally 
symmetric state, quantum fluctuations rapidly change it to a less coherent and more asymmetric one. 
In addition, due to (\ref {multysuinf}), irreducible representations of $SU(\infty)$ are partially 
entangled~\cite{targetspacebreak} and there is high probability of clustering of subspaces in a randomly 
selected~state. 

Lets assume that such groupings indeed have occurred in the early Universe and they continue to occur at Planck 
scale. They provide the necessary conditions for division of the Universe to parts or particles with 
$SU(\infty) \times G \cong SU(\infty)$ as their symmetry. The~local symmetry $G$ is assumed to be a 
compact Lie group of finite rank and respected by all subsystems. Although~different subsystems may have 
different internal symmetries, without lack of generality we can assume that $G$ is their tensor product, but~
some species of particles/subsystems are in singlet representation of some of the component~groups. 

As the rank of $G$ is assumed to be finite, complementarity condition dictates that the number of subsystems 
must be infinite to account for the infinite rank of $\hm_U$. If~states are in a finite dimensional 
representations of $G$, at~least one of the representations must have infinite multiplicity and its Hilbert 
space should be infinite dimensional. Thus, despite the division of $\bm [\hm_U]$, $SU(\infty)$ remains a symmetry 
of subsystems and $\{A_i\} \subset \bm [\hm_i] \cong SU(\infty)$, where $\hm_i$ is the Hilbert space of 
subsystem $i$. Clustering of states and subsystems are usually the hallmark of strong interaction and 
quantum correlation~\cite{anyonrev}. Therefore, the~interaction of subsystems through internal $G$ symmetry 
is expected to be stronger than through $SU(\infty)$, thereby the weak gravitation 
conjecture~\cite{weakgrconj} is~satisfied. 

We could also formulate the above Universe in a bottom-up manner. Consider the ensemble of infinite number of 
quantum systems---particles---each having finite symmetry $G$ and coherently mixed with each others. Their 
ensemble generates a Universe with $SU(\infty) \times G \cong SU(\infty)$ as symmetry represented by its 
Hilbert space. Therefore, top-down or bottom-up approaches to an infinitely divisible Universe give the 
same result. The~bottom-up view helps to better understand the origin of $SU(\infty)$ symmetry. It~shows that, for~each subsystem, it is the presence of other infinite number of subsystems and its own interaction 
with them that is seen as a $SU(\infty)$ symmetry.

\subsection{Properties of an Infinitely Divided Quantum~Universe} \label{sec:subuniveprop}
The division of the Universe to subsystems has several consequences. First of all, the~global $U(1)$ symmetry of 
$\hm_U$ becomes local, because~Hilbert spaces of subsystems $\hm_i,~\forall i$, where index $i$ runs over 
all subsystems, acquire their own phase symmetry. Therefore, we expect that there is at least one unbroken 
$U(1)$ local---gauge---symmetry in nature. It may be identified as $U(1)$ symmetry of the Standard Model. 
From now on, we include this $U(1)$ to the internal symmetry of subsystems $G$. Additionally, the~infinite 
number of subsystems in the Universe means that each of them has its own representation of $SU(\infty)$ 
symmetry. However, these representations are not isolated and are part of the $SU(\infty)$ symmetry of the 
whole Universe. This property is similar to finite intervals on a line, which are homomorphic to $R^{(1)}$ 
and, at~the same time, part of it and have the same algebra. Therefore, the~memory of being part of the 
whole Universe is not washed out with the division to subsystems. Otherwise, according to the corollary 
discussed in Appendix \ref{app:qmaxioms} subsystems could be considered as separate and isolated~universes. 

The area of diffeo-surface is irrelevant when only one $SU(\infty)$ is considered. However, it becomes relevant 
and observable when it is compared with its counterparts for other subsystems. More precisely, homomorphism 
between Hilbert spaces of two subsystems $s$ and $s'$ defined as: 
\be
\mathcal {R}_{ss'}: \hm_s \rightarrow \hm_{s'} \label{subsyshomo}
\ee
can be considered as an additional parameter that is necessary for their identification and indexing. A~more 
qualitative description of how a third continuous parameter emerges from division of Universe to subsystem 
is given in the next~subsection.

\subsection {Parameterization of Subsystems} \label {sec:radiusparam}
There are various ways to see that the division of the Universe to subsystems defined in Section~\ref{sec:infuniverse} 
induces a new continuous parameter. As~discussed in the previous subsection, each~subsystem represents
$SU(\infty) \times G$. When $SU(\infty)$ representation of different subsystems are compared, e.g.,~through a 
morphism, the~radius of diffeo-surface becomes relevant, because~different radius means different 
area. This dependence allows for classifying subsystems according to a size scale. More precisely, in~the 
definition of $\hL_{lm}$ in (\ref{lharminicexp}), $Y_{lm} \propto r^l$, where $r$ is the distance to centre 
in spherical coordinates when the 2D surface is embedded in $R^{(3)}$. If~we factorize $r$-dependence part 
of $Y_{lm}$, the~algebra of $\hL_{lm}$ defined in (\ref{lcommut}) becomes:
\be
[\hL_{lm},~\hL_{l'm'}]\biggl |_{r=1} = r^{l'' - l' - l} f ^{l"m"}_{lm,l'm'}~\hL_{l"m"}\biggl |_{r=1} \label{lcommutr}
\ee
where all $\hL_{lm}$ operators are defined for $r=1$ (in an arbitrary unit). Equation (\ref {lcommutr}) 
shows that $r$ can be interpreted as a coupling that quantifies the strength of correlation between $\hL_{lm}$ 
operators. Moreover, due to homomorphism (\ref {multysuinf}), $\hL_{lm}$'s of subsystems 
are part of $\hL_{lm}$'s of the full system. Consequently, subsystems are never completely isolated and they interact 
through an algebra similar to ({\ref{lcommutr}), but~their $r$ factors can be different:
\be
[\hL^{(r)}_{lm},~\hL^{(r')}_{l'm'}]\biggl |_{r=1} = {r``}^{l''} {r'}^{-l'} r^{-l} f ^{l"m"}_{lm,l'm'}~\hL^{r''}_{l"m"}
\biggl |_{r=1} \label{lcommutdiffr}
\ee
where $r$ indices on $\hL_{lm}$ operators are added to indicate that they may belong to different subsystems. 
Nonetheless, the~algebra remains the same, because~operators $\hL_{lm}$ belong, at~the same time, to~the global 
$SU(\infty)$. On~the other hand, the~nonlocality of this algebra in the point of view of subsystems should induce a dependence on derivative with respect to parameters when infinitesimal transformations are considered, e.g.,~
in the Lagrangian. Specifically, we expect a relation between $r''$ and $(r, r')$, determined by homomorphism 
(\ref {subsyshomo}). In~the infinitesimal limit, the~r.h.s. of (\ref{lcommutdiffr}) becomes Lie derivative 
of $\hL_{lm}$ in the direction of $\hL_{l'm'}$ in the manifold that is defined by parameters $(r, \theta, \phi, t)$, 
where the last parameter is time with respect to an observer, as~described in the next~subsection.

In summary, after~the division of the Universe to subsystems, their $SU(\infty)$ symmetries are indexed 
by angular parameters $(\theta, \phi)$ and an additional continuous parameter $r =(0, \infty)$. They share 
the algebra of global $SU(\infty)$, but~acquire a new index and, in~this sense, their algebra becomes 
nonlocal. Notably, in~the infinitesimal limit the algebra can be considered as the Lie derivative of 
$\hL_a \in \bm [\hm]$ operators on the manifold of parameter space $(r, \theta, \phi, t)$. Differential 
properties of the model need more investigation and will be reported~elsewhere. 

Finally, we can define a conjugate set of parameters for the dual Hilbert space $\hm_U^*$ and dual operators 
$\hJ_a$ defined in (\ref {lquantize}). Therefore, in~contrast to some quantum gravity candidates, this model does not have a preference for position or momentum~spaces.


\subsection{Clocks and~Dynamics} \label{sec:clock}
The last step for construction of a dynamical quantum Universe is the introduction of a clock by using 
comparison between variation of states of two subsystems, tagged as system and clock, under~the application of 
operators $\hL_\alpha \in SU(\infty) \times G$ by a third subsystem, tagged as observer, who~plays the 
role of a reference. The~necessity of an observer/reference is consistent with the foundation of quantum 
mechanics, as~described in~\cite{houriqmsymm}. In~the context of the present model, this discrimination can 
be understood as the following: although the global $SU(\infty)$ symmetry means that any variation of full 
state by application of $\hL_\alpha$ is a gauge transformation, a~variation of subsystems with respect to each 
others is meaningful and can be~quantified.

The technical details of introducing a clock and relative time in quantum mechanics are intensively studied in 
the literature, see e.g.,~\cite{qmtimedef} for a review and proof of the equivalence of different approaches. 
Here, we describe this procedure through an example. Consider the application of operators 
\mbox{$\hL_c \in \bm [\hm_C]$} and $\hL_s \in \bm [\hm_s]$ to two subsystems, called clock and system, 
respectively, such that:
\be
\hL_c \rho_c \hL_c^\dagger = \rho_c + d\rho_c \equiv \rho_c', \quad \quad \hL_s \rho_s \hL_s^\dagger = 
\rho_s + d\rho_s \equiv \rho'_s \label{diffrho}
\ee

Because these operations are local and restricted to subsystems, they are not gauged out. One way of 
associating a c-number quantity to these variations is to define parameter $t$, such that, for~instance,  
$dt \equiv |\tr~(\rho'_c \hO_c) - \tr~(\rho_c \hO_c)|$, where $\hO_c$ is an observable of the clock subsystem. 
This quantity is positive and, by~definition, incremental. The~Hamiltonian 
operator of the system $H_s \in \bm [\hm_s]$ according to this clock would be an operator for which 
$d\rho_s / dt = -i / \hbar [H_s, \rho_s]$. 

More generally, defining a clock is equivalent to comparing excursion path of two subsystems in their respective Hilbert space under successive application of $\hL_c$ and $\hL_s$ to them, respectively. The~
arrow of time arises because through the common $SU(\infty)$ symmetry any operation---even a local one---is communicated to the whole Universe. Thus, inverting the arrow of time amounts to performing an inverse 
operation on all subsystems, which is extremely difficult. Therefore, although~the dynamical equation of one 
system may be locally symmetric with respect to time reversal, due to global effect of every operation, its effect cannot be easily reversed.

\subsection {Geometry of Parameter Space} \label{sec:paramgeo}
The final stride of time definition brings the dimension of continuous parameter space necessary for 
describing states and dynamics of an infinite dimensional divisible Universe to 3 + 1, namely 
$(r, \theta, \phi, t)$.\footnote{Evidently, in~addition to 3 + 1 external parameters each subsystem 
represents the internal symmetry $G$, where its representations have their own parameters.} Although~these 
parameters arise from different properties of the Universe, namely $(\theta, \phi)$ from $SU(\infty)$ 
symmetry, $r$ from division to infinite number of subsystems, and~$t$ from their relative variation, they are 
mixed through the global $SU(\infty)$ symmetry, arbitrariness of the choice of reference frame and clock, and~
quantum superposition of states. Therefore, geometry of the parameter space is $R^{(3 + 1)}$. 

The 2D parameter space of the whole Universe is, by~definition, diffeomorphism invariant, as~it is the 
representation of $SU(\infty)$. However, at~this stage it is not clear whether the subdivided Universe is 
rigid, that is only invariant only under global frame transformations of the (3 + 1)D parameter space, or~
deformable and invariant under its diffeomorphism. Here we show that it is indeed diffeomorphism invariant. 
Moreover, its geometry is determined by states of subsystems.\footnote{Notice that even in classical general 
relativity diffeomorphism and relation between geometry and state of matter are independent concepts. 
In particular, Einstein equation is not the only possible relation and a priori other diffeomorphism 
invariant relations between geometry and matter are allowed---but constrained by experiments.} 

Consider a set of 2D diffeo-surfaces representing $SU(\infty)$ symmetries of subsystem. These diffeomorphism 
can be obtained from application of $\hL_{lm} \in SU(\infty)$ operators to vacuum state of each subsystem, 
considered to be a sphere. They are smooth functions of parameters $(r, \theta, \phi)$ and can be identified 
with states of the subsystems, which are also smooth functions of the same parameters. After~ordering these 
surfaces---for instance, according to their average distances \footnote{More generally, any measure of difference 
between states, such as Fubini--Study metric or fidelity can be used to order states. As~Hilbert spaces of 
quantum systems with $SU(\infty)$ symmetry consist of continuous functions, we can use usual analytical tools for 
defining a distance. However, we should not forget that functions are vectors of a Hilbert space. Moreover, 
Hilbert~space vectors are, in~general, complex functions and each projection between diffeo-surfaces 
corresponds to two projections in the Hilbert space, one for real part and one for imaginary part of vectors.}---and defining a projection between neighbours \footnote{This projection is isomorphic to a homomorphism 
between $\bm[\hm_s]$ of subsystems.}, such that if on $i\text{th}$ surface the point $(\theta_i, \phi_i, r_i)$ is 
projected to $(\theta_{i+1}, \phi_{i+1}, r_{i+1})$ on \mbox{$(i+1)\text{th}$} surface, the~distance in $R^{(3)}$ between 
points in an infinitesimal surface $\Delta \Omega_i < \epsilon^2$ containg $(\theta_i, \phi_i, r_i)$ and 
infinitesimal surface $\Delta \Omega_{i+1} < \epsilon'^2$ containg $(\theta_{i+1}, \phi_{i+1}, r_{i+1})$ 
approaches zero if $\epsilon, \epsilon'\rightarrow 0$. The~path connecting closest points on $\Delta \Omega_i$ 
and $\Delta \Omega_{i+1}$ defines an orthogonal direction in a deformed \mbox{$S_2 \times R^{(1) \cong }R^{(3)}$} and the Riemann curvature of this space can be determined from sectional curvature. Therefore, parameter space 
(or equivalently Hilbert space) is curved. Moreover, as~the projection between $\Delta \Omega_i$ and 
$\Delta \Omega_{i+1}$ used for this demonstration is arbitrary, we conclude that the parameter space is not 
rigid and its diffeomorphism does not change the physics. The~same procedure can be applied when a clock 
is chosen. Therefore, the~above conclusions apply to the full (3 + 1)D parameter space of the subdivided 
Universe.

Finally, from~homomorphism between diffeo-surfaces and states of subsystems, we conclude that (3 + 1)D classical 
spacetime can be interpreted as parameter space of the Hilbert space of subsystems of the Universe, 
and gravity as the interaction that is associated to $SU(\infty)$ symmetry.

\subsection{Metric~Signature} \label{sec:signature}
Up to now we indicated the dimension of spacetime---parameter space of $SU(\infty)$ symmetry---after 
subdivision of the Universe as 3 + 1. This implicitly means that we have considered a Lorentzian metric with 
negative signature. In~special and general relativity, the~signature of metric is dictated by observation 
of the constant speed of light in classical vacuum. Indeed, diffeomorphism invariance, Einstein 
equation, and~interpretation of gravity as curvature of spacetime are independent of signature of the 
spacetime~metric.

In quantum mechanics, Heisenberg uncertainty relation imposes Mandelstam--Tamm constraint~\cite{qmspeed} 
on the minimum time necessary for the transition of a quantum state $\rho_1$ to another perfectly 
distinguishable state $\rho_2$~\cite{qmspeedlimit}: 
\bea
&& \Delta t \geqslant \frac{\hbar}{\sqrt{2}}~\frac{\cos^{-1} A (\rho_1, \rho_2)}{\sqrt{Q(\rho_1, \hH)}}, 
\nonumber \\
&& A (\rho_1, \rho_2) \equiv \tr (\sqrt {\rho_1}\sqrt {\rho_2}), \quad \quad Q(\rho, \hH) \equiv 
\frac{1}{2} |\tr([\sqrt{\rho}, \hH]^2)| \label{aqdef}
\eea
where $\hH$ is the system's Hamiltonian. Consider $\rho_1$ as the state of Universe after selecting and 
separating an observer and a clock and $\rho_2$ as an infinitesimal variation of $\rho_1$, that is 
$\rho_2 = \rho_1 + d\rho_1$. We assume that the clock is chosen, such that, in~(\ref{aqdef}), minimum time is 
achieved. Subsequently, (\ref{aqdef}) becomes:
\be
Q(\hH, \rho_1) dt^2 = \tr (\sqrt{d\rho_1} \sqrt{d\rho_1}^\dagger) \equiv ds^2 \label{separation}
\ee
This equation is similar to a Riemann metric for a system at rest with respect to the chosen coordinates 
frame for the parameter space $(r, \theta, \phi, t)$. On~the other hand, the~r.h.s. of (\ref{separation}) 
only depends on the variation of state and is independent of the chosen frame for parameters. Therefore, 
$ds$ is similar to an infinitesimal separation. A~coordinate transformation, i.e.,~
$(r, \theta, \phi, t) \rightarrow (r', \theta', \phi', t')$ does not change state of the Universe and is 
equivalent to a basis transformation in the Hilbert space. On~the other hand, the~l.h.s. of (\ref{separation}) 
changes. Considering the similarity of (\ref{separation}) to metric equation, we can write $ds^2$ as:
\be
g_{00} d{t'}^2 \pm g_{ii} d{x'}^i d{x'}^i = ds^2 \label{newmetric}
\ee
where we have used Cartesian coordinates in place of spherical. We have chosen parameter transformation such 
that $g_{0i} = g_{i0} = 0$. We have also assumed $g_{ij} > 0$ and factorized the sign of spatial part of the 
metric. In~these new coordinates, the~Hamiltonian associated to the new clock $t'$ is $\hH'$ and Mandelstam--Tamm 
relation imposes:
\be
Q'(\hH', \rho_1) d{t'}^2 \equiv g_{00} d{t'}^2 \geqslant \tr (\sqrt{d\rho_1} \sqrt{d\rho_1}^\dagger) = ds^2 
\label{separationp}
\ee

For $ds^2 \geqslant 0$, constraint (\ref{separationp}) is only satisfied if the sign of spatial part in 
(\ref{newmetric}) and thereby the signature of the metric is negative. We remind that Mandelstam--Tamm constraint 
does not apply to states that do not fulfill distinguishability condition. In~these cases, $ds^2 < 0$ is 
allowed. In~classical view of spacetime, they correspond to spacelike events, where two events/states are not causally related. In~quantum mechanics, this can be related to nonlocality~\cite{hiesenbernonlocal} 
and the absence of strict causality. Additionally, the~explicit dependence of separation on the density matrix 
in (\ref {separation}) and (\ref{separationp}) and its independence of the coordinate frame of the 
parameters/spacetime confirms and completes the discussion of Section~\ref{sec:paramgeo} regarding the curved geometry 
of parameter space and diffeomorphism invariance of subdivided~Universe.

\subsection{Lagrangian of~Subsystems}  \label{sec:sublagrangian}
Finally, the~Lagrangian of the Universe after the division to subsystems and selection of reference observer 
and clock takes the following form:
\bea
\mathcal {L}_{U_s} & = & \int d^4 x \sqrt {-g}\biggl [\frac{1}{16\pi G_N \hbar} \sum_{l,m,l',m'} \tr (L^*_{lm} (x)
 L_{l'm'}(x) \hL_{lm} \hL_{l'm'}) + \nonumber \\
&& \frac {1}{8 (\pi G_N \hbar)^{1/2}} \biggl (\sum_{l,m,a} \tr (L_{lm}(x) T_a (x) \hL_{lm} \otimes \hT_a) + 
\sum_{lm} L_{lm} \tr (\hL_{lm} \otimes {\mathbbm 1}_G \rho (x)) \biggr ) + \nonumber \\
&& \frac {1}{4} \sum_{a,b} \tr (T^*_a (x) T_b (x) \hT_a \hT_b) + 
\frac{1}{2} \sum_a T_a \tr ({\mathbbm 1}_{SU(\infty)} \otimes \hT_a \rho (x)) \biggr ]. \label{lagrang}
\eea

The terms of this Lagrangian can be interpreted as the following. The~first term is the Lagrangian for 
an ensemble of $SU(\infty)$ symmetries of all subsystems, except~observer and clock. Amplitudes $L_{l'm'}(x)$ 
depend on full $SU(\infty)$ parameter space, which is $(r, \theta, \phi, t)$. Due to the nonlocal algebra 
(\ref{lcommutdiffr}), we expect that $L_{l'm'}(x)$'s include derivative terms. Additionally, $L_{lm}$'s are 
normalized such that the usual gravitational coupling be explicit. We notice that, if~
$\hbar G_N ~ \propto ~ \hbar^2/M_P^2 \rightarrow 0$, the~first and the third terms will be canceled. Therefore, 
the~naive classical limit of the model does not include these gravity related terms. 

The second term presents gravitational interaction of internal gauge fields and matter, respectively. The third 
and forth terms together correspond to the Lagrangian of pure gauge fields for local $G$ symmetry and its 
interaction with matter field. They take the standard form of Yang--Mills models if $T_a (x)$ fields 
are two-forms in the (3 + 1)D parameter space. 

We leave explicit description of $L_{l'm'}(x)$'s and $T_a$'s as functionals of spacetime, and determination 
of semi-classical limit of the Lagrangian for future works. Nonetheless, the~Lagrangian (\ref{lagrang}) 
is not completely abstract. $L_{lm}$ operators can be expressed as a tensor product of Pauli matrices and 
regrouped by $r$ and $t$ indices, which have no other role than associating a group of matrices to 
subsystems. This is because the tensor product of $SU(\infty)$ is homomorphic to itself. However, such 
expansion is not very useful and practical for analytical calculations, in~particular for finding semi-classical 
limit of the~model.

\section{Comparison with Other Quantum Gravity~Models} \label{sec:compmodel}
It is useful to compare this model with string theory and Loop Quantum Gravity (LQG)---the two most popular 
quantum gravity candidates.

A common aspect of string/superstring theories with the present model is the 
presence of a 2D manifold in their foundation. However, in~contrast to string theories, in~which a 2D world 
sheet is introduced as an axiom without any observational support, the~presence of a 2D manifold here is a 
consequence of the infinite symmetry of the Universe, which has compelling observational support. Moreover, 
the 2D nature of the underlying Universe manifests itself only when the Universe is considered as a whole. 
Otherwise, it is always perceived as a (3 + 1)D continuum (plus parameters of internal symmetry of 
subsystems). 

In string theory, matter and spacetime are fields living on the 2D world sheet, or~equivalently the world 
sheet can be viewed as being embedded in a multi-dimensional, partially compactified space without any 
explanation for the origin of such non-trivial structures. On~the other hand, in the present model 
the approach to matter is rather bottom-up. The~Cartan decomposition of $SU (\infty)$ to smaller groups, 
in~particular $SU(2)$ means that they can easily break and separate from the pool of the $SU(\infty)$ 
symmetry---for instance by quantum correlation between pairs of subsystems---without affecting the infinite 
symmetry. And~indeed it seems 
to be the case because $SU(2)$ and $SU(3) \subset SU(2) \times SU(2) \cong SU(4)$ are Standard Model 
symmetries. Additionally, string theory is fundamentally first quantized and string based field theories 
are considered to be low energy effective descriptions. However, as~explained in the previous sections, 
in~the present model owing to its infinite dimensional symmetry, Hilbert and Fock spaces are homomorphic and 
the model can be straightforwardly considered as first or second~quantized.

The importance of $SU(2)$ symmetry in the construction of LQG and its presentation as spin foam~\cite{lqgfoam} 
is shared with the present model. However, $SU(2) \cong SO(3)$ manifold on which Ashtekar variables are 
defined has its origin in the ADM (3 + 1)D formalism, based on the presumption that spacetime and 
thereby quantum gravity should be formulated in the physical spacetime. Moreover, LQG does not address 
the origin of matter as the source of gravity or the origin of the Standard Model symmetries. 
The present model explains both the dimension of spacetime and relation between quantum gravity, matter, 
and SM~symmetries.

A concept that string theory and LQG do not consider---at least not in their foundation---is the fact 
that in quantum mechanics discrimination between observer and observered is essential, and models which 
do not consider this concept in their construction---especially when the models is intended to be applied to 
the whole Universe---are somehow metaphysical, because~they implicitly consider that the observer is out of 
this~Universe.

\section{Outline and Future~Perspectives} \label{sec:outline}
In this work, we proposed a new approach to quantum gravity by constructing a Universe in which gravity is 
fundamentally quantic and demonstrated how it may answer some of questions that we raised in the Introduction 
section regarding gravity and the nature of spacetime. As~we have already summarized the model and its results 
in Section~\ref{sec:summary}, here we concentrate on perspectives for further~studies. 

Understanding nonlocality and differential form of the algebra of subsystem defined by 
Equation~(\ref {lcommutdiffr}) is crucial for finding an algebraic expression for the Lagrangian (\ref{lagrang}), 
which, at~present, is~too abstract. This task is especially important for investigating the semi-classical limit 
of the model. On~the other hand, this Lagrangian describes an open system, because~the state of the observer and 
probably some of degrees of freedom of the clock are traced out. Formulation of the subdivided Universe as an 
open system should help application of the model to black hole physics and~cosmology.

We discussed a bottom-up procedure for the emergence of internal symmetries in Section~\ref{sec:division}. In~
particular, we concluded that they should generate stronger couplings between particles/subsystems than 
gravity. However, this argument does not explain how the hierarchy of couplings arises. We conjecture that 
clustering of subsystems, which leads to the emergence of internal symmetries, also determines their couplings, 
probably through processes that are analogous to the formation of moir\'e super-lattice and strong correlation 
between electrons in 2D materials. The~fact that, in~this model, both the Universe as a whole and its subsystems 
have $SU(\infty)$ symmetry, which is represented by diffeomorphism of 2D surfaces, means that the necessary 
ingredients for formation of moir\'e-like structures are readily~available.

In the absence of experimental quantum gravity tests, the~ability of models to solve theoretical issues has 
prominent importance. Among~topics that must be addressed black holes and puzzles of information loss in 
semi-classical approaches have high priority. Because~the model studied here is inherently quantic, the~first task 
is finding a purely quantic definition for black holes. Naively, a~quantum black hole may be defined as a 
many particle system in a quantum well in real space. However, we know that quantum field theory in curved 
spacetime background of black holes leads to Hawking radiation and extraction of energy from black hole. 
Consequently, in~the realm of quantum mechanics, black holes are not really contained in a limited region of 
space. Their potential well is not perfect and their matter content extends to~infinity. Thus, this definition 
should be considered as an initial condition.

Inflation and dark energy are other issues that should be investigated in the context of this model. 
Notably, it would be interesting to see whether the topological nature of 2D Lagrangian of the 
whole Universe can have observable consequences, for~instance, as~a small but nonzero vacuum energy.
As for inflation, an~exponential decoupling and decoherence of particles/subsystems in the early universe 
may be interpreted as inflation and an extension of spacetime. These possibilities need detailed~investigation.

In conclusion, the~inhomogeneous Lorentz transformation may be the classical interface of a much deeper 
and global realm of a quantum~Universe.

\paragraph*{\bf acknowledgments} The author thanks Institut Henry Poincar\'e for hospitality and bibliographic 
assistance during accomplishment of this~work.

\appendix
\section {A Very Brief Summary of the Best Studied Quantum Gravity Models} \label{app:qgrsumm} 
Introduction of quantum mechanical concepts to general relativity was first mentioned by Einstein 
himself in his famous 1916 paper. The~first detailed work on the topic was by L\'eon Rosenfeld in 
1930~\cite{qgrlagrangian}, in~which the action of Einstein-Hilbert model with matter is quantized by replacing 
classical variables with hermitian operators, see e.g.,~\cite{qgrearlyhist} for the history of early approaches 
to quantum gravity. This canonical approach and its modern variants based on the quantization of 3 + 1 dimensional 
Hamiltonian description of dynamics, notably Wheeler-DeWitt (WD) formalism~\cite{qgrcanonical,qgrcanonical0} 
and quantum geometrodynamics~\cite{qgrgeometrody} lead to nonrenormalizable models.\footnote{We should emphasize 
that references given in this appendix are only examples of works on the subjects on which tens or even 
hundreds of articles can be found in the literature.} See e.g.,~\cite{qgrgeometrodrev} for review of other 
issues of these approaches and their current~status.

Another model, inspired by the ADM Hamiltonian formulation of general relativity~\cite{admgr}, the~Dirac Hamiltonian 
description of quantum mechanics~\cite{qmhamilton}, and~the WD approach to QGR is Loop Quantum Gravity (LQG), 
see e.g.,~\cite{lqgrev,lqgrev0} and references therein. In~this approach, triads defined on a patch of the 
3D space---what is called Ashtekar variables~\cite{ashtekarvar}---replace spatial coordinates and are
considered as Hermitian operators acting on the Hilbert space of the Universe. Their conjugate operators
form a $SU(2)$ Yang-Mills theory and provide a connection---up to an undefined constant called Immirzi
parameter---for the quantized 3D space. However, to~implement diffeomorphism of general relativity without
referring to a fixed background, the~physical quantized entities are holonomies---gauge invariant
nonlocal fluxes and Wilson loops defined on 2D surfaces and their boundaries, respectively. Similar to the WD
formalism, the LQG Hamiltonian is a constraint, and~thereby there is no explicit time in the
model~\cite{lqgrev}. Recently, it is shown that a conformal version of the LQG has an explicit time
parameter~\cite{lqgconform}. But, conformal symmetry must be ultimately broken to induce a mass or distance
scale in the model. Other~issues in the LQG are lack of explicit global Lorentz invariance, absence of any
direct connection to matter, and~most importantly quantization of space, that violates Lorentz invariance
even when the absence of time parameter in the model is~neglected. 

Regrading the violation of Lorentz invariance, even if discretization is restricted to distances close 
to the Planck scale, matter interaction propagates it to larger distances~\cite{lqglorentz}. 
This issue is also present in other background independent approaches to quantum gravity, in~which in one 
way or another the spacetime is discretized. Examples of such models are symplectic quantum 
geometry~\cite{qgrsymplect} and dynamical triangulations, in~which space is assumed to consist of a 
dynamical lattice~\cite{qgrdyntrian,qgrdyntrian0}. See also~\cite{qgrdyntrian1} for a recent review of 
these approaches and~\cite{qgrdyntrian2} for some of their issues, in~particular a 
likely absence of a UV fixed point, which is necessary for renormalizabilty of these models. Therefore, the~
claimed quantization of space volume or in other words emergence of a fundamental length scale in UV limit 
of these models is still uncertain. Another example is causal sets---a discretization approach with causally 
ordered structures~\cite{qgrcausalset}, see e.g.,~\cite{qgrcausalset1} for a review. They probably suffer from 
the same issue as other discretization models, notably breaking of Lorentz symmetry, 
see e.g.,~\cite{qgrcausalsetsymmbreak}, but~also~\cite{qgrcausalsetsymm} for counter-arguments. We should 
remind that all quantum gravity models depend on a length (or~equivalently mass) scale, namely the Planck 
length $L_P$ (or mass $M_P$). Dimensionful quantities need a {\it unit}, which does not arise from 
dimensionless or scale invariant quantities. Therefore, discretization is not a replacement for a 
dimensionful fundamental constant in quantum gravity~models.

Another way of quantizing spacetime without discretization is consideration of a noncommutative 
spacetime~\cite{noncummut,noncummut0}. This formalism is in fact one of the earliest proposals for a quantum 
gravity. More~recently this approach is studied in conjunction with other QGR models such as string 
theory~\cite{noncommutstring} and matrix models~\cite{noncommutmatrix}. An~essential issue of this class 
of models is their inherent nonlocality that leads to mixing of low and high energy 
scales~\cite{noncommutuvirmix}. On~the other hand, this characteristic might be useful for constraining 
them, and~thereby related QGR models~\cite{noncommutcmbconst}.

In early 1980's the discovery of both spin-1 and spin-2 fields in 2D conformal quantum field theories 
embedded in a $D$-dimensional spacetime---called {\it string} models---opened a new era and discipline for 
seeking a reliable quantum model for gravity, and~ultimately unifying all fundamental forces in a 
{\it Great Unified Theory (GUT)}. Nambu-Goto and Polyakov string theories were studied in 
1970's as candidates for describing strong interaction of hadrons. Although~with the establishment of 
Quantum Chromo-Dynamics (QCD) as the true description of strong nuclear force string theories seemed 
irrelevant, their potential for quantizing spacetime~\cite{qmgeostring,qmgeostring0} gave them a new role 
in fundamental particle physics. String and superstring theories became and continue to be by far the most 
extensively studied candidates of quantum gravity and GUT.\footnote{A textbook description and references 
to original works can be found in textbooks such as~\cite{stringrev,stringrev0}.}

Quantized strings/superstring models are finite and meaningful only for special values of spacetime 
dimension $D$. For~these cases, the~central charge of Virasoro algebra or its generalization to affine Lie 
algebra vanishes when the contribution of all fields, including ghosts of the conformal theory on the 2D 
world-sheet are taken into account. Without~this restriction the theory is infested by anomalies, singularities, 
and misbehaviour. The~allowed dimension is $D=26$ for bosonic string theories and $D=10$ for superstrings. 
The group manifold on which a viable string model can live is restricted as well. For~instance, the~allowed 
symmetry in heterotic Polyakov model is $SO(32)$ or $E_8 \times E_8$. Wess-Zumino-Novikov-Witten (WZNW) models 
with 2D affine Lie algebras provide more variety of symmetries, including coset groups. However, restriction 
on dimension/rank of symmetry groups remains the same. Therefore, to~make contact with real world, which 
has 3 + 1 dimensions, the~remaining dimensions must be~compactified. 

Initially the inevitable compactification of fields in string models was welcomed because it might explain 
internal global and local (gauge) symmetries of elementary particles, in~a similar manner as in Kaluza-Klein 
unification of gravity and electromagnetism~\cite{kaluza,klein}. However, intensive investigations of the 
topic showed that compactification generates a plethora of possible models. Some of these models may be 
considered more realistic than others based on the criteria of having a low energy limit containing the 
Standard Model symmetries. But, unobserved massless moduli, which may make the Universe overdense if they 
acquire a mass at string or even lower scales, strongly constrain many of string models. Therefore, moduli 
must be stabilized~\cite{moduliproblem,modulistab}. For~instance, they should acquire just enough effective 
mass to make them a good candidate for dark matter~\cite{dmmoduli}. Moreover, in~string theories there is no 
natural inflation candidate satisfying cosmological observations without fine-tuning. Although~moduli are 
considered as potential candidates for inflation~\cite{modulieinf}, small non-Gaussianity of Cosmic Microwave 
Background (CMB) anisotropies~\cite{plancknongauss} seems to prefer single field 
inflation~\cite{infsinglenongauss}. In~addition, single field slow roll inflation may be 
inconsistent~\cite{infslowrollswap} with constraints to be 
imposed on a scalar field interacting with quantum gravity in the framework of swampland extension of string 
models landscape~\cite{stringswampland}. Some researchers still believe that a genuinely non-perturbative 
formulation of superstring theories may solve many of these issues \footnote{Non-supersymmetric string 
models may have no non-perturbative formulation and should be considered as part of a supersymmetric 
model, see e.g.,~Chapter 8 of~\cite{stringrev0}.}. However, the~absence of any evidence of supersymmetry 
up to $\sim$TeV energies at LHC---where it was expected, such that it could solve Higgs hierarchy 
problem~\cite{higgshier}---is another disappointing result for string~models.

Observation of accelerating expansion of the Universe due to a mysterious dark energy with properties very 
similar to a cosmological constant---presumably a nonzero but very small vacuum energy---seems to be another 
big obstacle for string theory~\cite{stringvacuarev} as the only quantum gravity candidate including both 
matter and gravity in its construction. The~landscape of string vacua has $\gtrsim$$10^{200-500}$~minima---depending on how models are counted~\cite{landscapecount}. But~there is no rule to determine which one is more 
likely and why the observed density of dark energy---if it is the vacuum energy---is $\sim$$10^{123}$ fold less 
than its expected value, namely ${M_P}^4$. To~tackle and solve some of these issues, extensions and/or 
reformulations of string theories have led to their variants such as matrix models~\cite{qgrmatrix,qgrmatrix0}, 
M-theory, F-theory, and~more recently swampland~\cite{stringswampland} and weak gravity 
conjecture~\cite{weakgrconj,weakgrbh}, and~models constructed based on~them.

In early 1999 Randall-Sundrum brane models~\cite{rs,rs0} and their variants---inspired by D-branes in toroidal 
compactification of open strings and propagation of graviton closed strings in the bulk of one or two  
non-compactified warped extra dimensions---generated a great amount of excitement and were subject of intensive 
investigations. By~confining all fields except gravitons on 4D branes these models are able to lower the 
fundamental scale of quantum gravity to TeV energies---presumably the scale of weak interaction---and explain 
the apparent weakness of gravitational coupling and high value of Planck mass. Thus, a~priori brane models 
solve the problem of coupling hierarchy in Standard Model of particle physics. In~addition, an~effective small 
cosmological constant on the visible brane may be achievable~\cite{branevacuum,branevacuum0}. 
However, brane models, in~general, have a modified Friedmann equation, which is strongly constrained by 
observations~\cite{braneex,braneex0,braneex1}. Moreover, it is shown that the confinement of gauge bosons on 
the brane(s) violates gauge symmetries, and~if gauge fields propagate to the bulk, so do the 
matter~\cite{rubakovvect,rubakovvect0}. Nonetheless, some methods for their localization on the brane 
are suggested~\cite{rubakovvect1,branevectlocal}. On~the other hand, observation of ultra high energy cosmic 
rays constrains the scale of quantum gravity and characteristic scale of warped extra-dimension to 
$>$100~TeV~\cite{houribraneqcd,houribraneqcd0}. This constraint is consistent with other theoretical and 
experimental issues of brane models, specially in the context of black hole physics, that is instability of 
macroscopic black holes, nonexistence of an asymptotically Minkowski solution~\cite{branebh,branebh0}, and~
observational constraint~\cite{branebhlhcex} on the formation of microscopic black holes in colliders at TeV 
energies~\cite{branebhlhc}.

In the view of these difficulties more drastic ideas have emerged. Some authors suggests UV/IR 
correspondence of gravity. They propose that at UV scales graviton quantum condensate behaves asymptotically 
similar to classical gravity~\cite{grsemiclass,grsemiclass0}. Other proposals attracting some interest 
include the emergence of classical gravity and spacetime from thermodynamics and 
entropy~\cite{grthermo,grthermo0} or condensation of more fundamental 
fields~\cite{qgremergecond,qgremergecond0}. 

More recently, the~development of quantum information theory and its close relation with entanglement of 
quantum systems, their entropy and the puzzle of information loss in Hawking radiation of black holes 
have promoted models that interpret gravity and spacetime as an emergent effect of 
entanglement~\cite{qgrautomatom,qgrentangle,qgrentangle0} and tensor networks~\cite{qgrtensornet,qgrtensornet0}. 
These ideas are in one way or another related to holography principle and Ads/CFT equivalence 
conjecture~\cite{adscft}. In~these models spacetime metric and geometry emerge from tensor decomposition of 
the Hilbert space of the Universe to entangled subspaces. The~resulted structures are interpreted as graphs 
and a symplectic geometry is associated to them. In~the continuum limit the space of graphs can be considered 
as a quantum spacetime. In~a somehow different approach in the same category of models the concept of 
{\it locality} specified by subalgebras is used to decompose the Universe. Local observables belong to 
spacelike subspaces in a given reference frame/basis~\cite{qgrlocalqm,qgrlocalqm0}. This means that in these 
models a background spacetime is implicitly postulated without being precise about its origin and nature. 
In addition to spacetime, subsystems/subalgebras should somehow present matter. But, it is not clear how 
they are related. Moreover, the~problem of the spacetime dimensionality and how it acquires its observed 
value is not discussed. In~any case, investigation of these approaches to quantum gravity is still in its 
infancy and their theoretical and observational consistency are not fully worked~out.

\section{Quantum Mechanics Postulates in Symmetry~Language} \label{app:qmaxioms} 
In this appendix we reformulate axioms of quantum mechanics \`a la Dirac~\cite{qmdirac} and 
von Neumann~\cite{qmvonneumann} with symmetry as a foundational concept:
\setcounter{enumi}{0}
\renewcommand{\theenumi}{\roman{enumi}}
\begin{enumerate}
\item A quantum system is defined by its symmetries. Its state is a vector belonging to a projective 
vector space called {\it state space} representing its symmetry group. Observables are associated 
to self-adjoint operators. The~set of independent observables is isomorphic to subspace of 
commuting elements of the space of self-adjoint (Hermitian) operators acting on the state space and 
generates the maximal abelian subalgebra of the algebra associated to symmetry group. 
\label{poststate}
\item The state space of a composite system is homomorphic to the direct product of state spaces of 
its components.\footnote{Notice that this axiom differentiates between possible states of a composite system, 
which is the direct product of those of subsystems, and~what is actually realized, which can be limited to a 
subspace of the direct product of individual components and have reduced symmetry.} In~the special case of 
separable components, this homomorphism becomes an isomorphism. Components may be separable-untangled---in 
some symmetries and inseparable---entangled---in others. The~symmetry group of the states of a composite 
system is a subgroup of direct product of its components. 
\label{postcomposite}
\item Evolution of a system is unitary and is ruled by conservation laws imposed by its symmetries and 
their representation by the state space. \label{postunitary}
\item Decomposition coefficients of a state to eigen vectors \footnote{More precisely {\it rays} because 
state vectors differing by a constant are equivalent.} of an observable presents the coherence/degeneracy 
of the system with respect to its environment according to that observable. Projective measurements 
{\bf by definition} 
correspond to complete breaking of coherence/degeneracy. The~outcome of such measurements 
is the eigen value of the eigen state to which the symmetry is broken. This spontaneous decoherence 
(symmetry breaking) \footnote{Ref.~\cite{houriqmsymm} explains why decoherence should be considered as a 
spontaneous symmetry breaking similar to a phase transition.} reduces the state space to the subspace 
generated by other independent observables, which represent remaining symmetries/degeneracies.\label{postmeasure}
\item A probability independent of measurement details is associated to eigen values of an 
observable as the outcome of a measurement. It presents the amount of coherence/degeneracy 
of the state before its breaking by a projective measurement. Physical processes that determine 
the probability of each outcome are collectively called {\it preparation}.\footnote{Literature on the 
foundation of quantum mechanics consider an intermediate step called {\it transition} between preparation and 
measurement. Here we include this step to preparation or measurement operations and do not consider 
it as a separate physical operation.} \label{postsymmbr}
\end{enumerate}

These axioms are very similar to their analogues in the standard quantum mechanics, except~that we do not 
assume an abstract Hilbert space. The~Born rule and classification of the state space as a Hilbert space 
can be demonstrated using axioms (\ref{poststate}) and (\ref{postsymmbr}), and~properties of statistical 
distributions~\cite{houriqmsymm}. We remind that the symmetry represented by the Hilbert space of a quantum 
system is in addition to the global $U(1)$ symmetry of states, which leaves probability of outcomes in a 
projective measurement unchanged. When system is divided to subsystems that can be approximately considered 
as non-interacting, each subsystem acquire its own {local} $U(1)$ symmetry. Even in presence of interaction 
between subsystems, a~local $U(1)$ symmetry can be considered, as~long as the interaction does not change the 
Hilbert space of subsystems. We notice that axiom~(\ref{postcomposite}) slightly diverges from its analogue in 
the standard quantum mechanics. It emphasizes on the fact that the symmetry group represented by a composite 
system can be smaller than the tensor product of those of its components. In~particular, entanglement may 
reduce the dimension of Hilbert space and thereby the rank of symmetry group that it~represents. 

A corollary of these axioms is that without division of the Universe to {\it system(s)} and {\it observer(s)} 
the process of measurement is meaningless. In~another word, an~indivisible universe is trivial and 
homomorphic to an empty set. In~standard quantum mechanics the necessity of the division of the Universe 
to subsystems arises in the Copenhagen interpretation, which has many issues, see e.g.,~\cite{copenhagenissue} 
for a review. In~covariant quantum models and ADM canonical quantization of gravity, in~which Hamiltonian 
is always null and naively the Universe seems to be static, relational definitions of time is based on the 
division of the Universe to subsystems, see e.g.,~\cite{qmtimedef}. Therefore, we conclude that division to 
subsystems is fundamental concept and must be explicitly included in the construction of quantum cosmology 
models.

\section{State Space Symmetry and~Coherence} \label{app:cohersymm}
The choice of a Hilbert space $\hm$ to present possible states of a system is usually based on the symmetries 
of its classical Lagrangian. Although~these symmetries have usually a finite rank---the number of 
simultaneously measurable observables---the Hilbert space presenting them may be infinite dimensional. For~
example, translation symmetry in a 3D space is homomorphic to $U(1) \times U(1) \times U(1)$ and has a global 
$SU(2) \cong SO(3)$ symmetry under rotation of coordinates. They can be presented by 6 parameters/observables. 
Thus, the~rank of the symmetry is finite. Nonetheless, due to the abelian nature of $U(1)$ group, 
the Hilbert space of position operator $\hm_X$ is infinite dimensional. More~generally, the~dimension of the 
Hilbert space depends on the dimension of the representation of the symmetry group of Lagrangian and its 
reducibility. The~Hilbert space of a multi-particle system can be considered as a reducible representation of 
the symmetry, even if single particles are in an irreducible representation. In~particular, Fock space of 
a many-particle system can be presented as an infinite dimensional Hilbert space representing symmetries 
of the Lagrangian in a reducible manner. This property is important for the construction of the quantum 
Universe model studied here, because~it demonstrates that the infinite size of the physical space can be 
equally interpreted as manifestation of infinite number of particles/subsystems in a composite~Universe.

Ensemble of linear operators acting on a Hilbert space $\bm[\hm]$ represents $SU(N)$ group where $N$ is the 
dimension of the Hilbert space $\hm$ and can be infinite. As~discussed in details in~\cite{houriqmsymm} 
configuration space of classical (statistical) systems have $\bigotimes^N U(1)$ symmetry where each $U(1)$ 
is isomorphic to the continuous range of values that an observable may have. Thus, quantization extends the 
symmetry of classical configuration space to $\bm[\hm] = SU(N) x U(1) \cong U(N) \supset \bigotimes^N U(1)$, 
where here we have also considered the global $U(1)$ symmetry of the Hilbert~space. 

Application of linear operators can be interpreted as interaction with another system or more generally with 
the rest of the Universe. The~change of state can be also considered as Positive Operator Valued Measurement 
(POVM). 
In particular, a~projective measurement and decoherence makes the state completely 
incoherent $\hat{\rho}_{inc}$:
\be
\hat{B} \hat{\rho}_c \hat{B}^\dagger \rightarrow \hat{\rho}_{inc} = 
\sum_i \rho_i \hat{\rho}_i, \quad \quad \hat{\rho}_i \equiv |i\rangle\langle i| \label{incoherestate}
\ee
where $\hat{B} \in \bm[\hm]$, $|i\rangle$ is an eigen basis for the measured observable, and~subscript 
$_{inc}$ means incoherent. We remind that the space of simultaneously observable operators corresponds to 
the Cartan subalgebra of $\bm[\hm]$. Coefficients $\rho_i$ are probability of occurrence of eigen value 
of $|i\rangle$ as outcome of the measurement. Because~$ \hat{\rho}_{inc}$ is diagonal, completely 
incoherent states $\hat{\rho}_{inc}$ represent the Cartan subgroup of $\bm[\hm]$. A~maximally coherent 
state in the above basis is defined as:
\be
\hat{\rho}_{maxc} \propto \sum_{i,j} |i\rangle\langle j| \label{maxcoherestate}
\ee
This is a pure state, in which all eigen states have the same occurrence probability in a projective 
measurement. Notice that due to the projectivity of Hilbert space $\hat{\rho}_{maxc}$ is unique and 
application of any other member of $\bm[\hm]$ reduces its coherence, quantified for instance by 
fidelity or Fubini-Study metric~\cite{qminfocohere}. More generally, action of $\bm[\hm]$ members changes 
coherence of any state which is not completely incoherent. For~this reason, we call $SU(N)$ symmetry of 
$\bm[\hm]$ the {\it coherence symmetry}.\footnote{In some quantum information literature coherence symmetry 
is called {\it asymmetry}~\cite{qmspeedlimit}. In~this work we call it {\it coherence symmetry} or simply 
{\it coherence} to remind that its origin is quantum degeneracy and indistinguishability/symmetry of states 
before a projective observation.}

It is useful to remind that in particle physics generators of $\bm[\hm]$ space physically exist and are not 
abstract operation of an apparatus controlled by an experimenter. In~the Standard Model $\bm[\hm]$ is 
generated by vector boson gauge fields in fundamental representation of SM symmetry group. They act on the 
Hilbert space generated by matter fields. If~gravity, which is the only known universal interaction follows 
the same rule, we should be able to define a Hilbert space for matter on which linear operators representing 
gravity~act. In the model studied here we identify these operators with $\hL$'s defined in Sec. \ref {sec:infuniverse}.

Regarding the example of translational and rotational symmetries of the physical space mentioned earlier, 
despite the fact that the dimension of the Cartan subalgebra of $\bm[\hm_X] \cong SU(N \rightarrow \infty)$ 
is infinite, and~a priori there must be infinite simultaneously observable quantities in the physical space, 
in quantum mechanics only one vector observable is associated to $\bm[\hm_X]$, namely the position of a 
particle/system. QFTs define field operators at every point of the space and assume that at equal time 
operators at different positions commute (or in the case of fermions anti-commute). However, in~the 
formulation of QFT models position is a parameter not an operator. These different interpretations of 
spacetime highlight the ambiguity of its nature in quantum contexts---as described in question 
\ref{matterspace} in the Introduction~section.

\section{\boldmath{$SU(\infty)$} and Its Polynomial Representation}  \label{app:harmonicdecomp}
Special unitary group $SU(N)$ can be considered as $N$-dimensional representation of $SU(2)$. For~this reason 
generators $T^{(N)}_{lm}$ of the associated Lie algebra $su(N)$ can be expanded as a matrix polynomial of 
$N$-dimensional generators of $SU(2)$. Indices $(l,m)$ in these generators are the same as in $SU(2)$ 
representations: $l=0,\cdots,N-1,~m=-l \ldots,+l$. Lie bracket of generators $T^{(N)}_{lm}$ is defined as:
\be
[\hT^{(N)}_{lm},~\hT^{(N)}_{l'm'}] = f^{(N)~l"m"}_{lm,l'm'}~\hT^{(N)}_{l"m"} \label{tcimmut}
\ee
Structure coefficients $f ^{(N)~l"m"}_{lm,l'm'}$ of $su(N)$ can be written with respect to 3j and 6j symbols, 
see e.g.,~\cite{suninfhoppthesis} for their explicit expression. For~$N\rightarrow \infty$, after~rescaling 
these generators $\hT^{(N)}_{lm} \rightarrow (N/i)^{1/2} \hT^{(N)}_{lm}$, they~satisfy the following Lie brackets:
\be
[\hL_{lm},~\hL_{l'm'}] = f ^{l"m"}_{lm,l'm'}~\hL_{l"m"} \label{lcommut}
\ee
where $\hL_{lm} \equiv \hT^{(N\rightarrow \infty)}_{lm}$ and coefficients $f ^{l"m"}_{lm,l'm'}$ are $N\rightarrow \infty$ 
limit of $f ^{(N)~l"m"}_{lm,l'm'}$. In~addition, it is shown~\cite{suninfhoppthesis} that $\hL_{lm}$ can be 
expanded with respect to spherical harmonic functions $Y_{lm}(\theta, \phi)$ defined on a sphere, i.e.,~the~manifold associated to $SU(2)$:
\bea
&& \hL_{lm} = \frac{\partial Y_{lm}}{\partial \cos \theta} \frac{\partial}{\partial \phi} - 
\frac{\partial Y_{lm}}{\partial \phi} \frac{\partial}{\partial \cos \theta} \label{lharminicexp} \\
&& \hL_{lm} Y_{l'm'} = -\{Y_{lm},~Y_{l'm'}\} = -f ^{l"m"}_{lm,l'm'} Y_{l"m"} \label{lapp} \\
&& \{\mathsf{f},~\mathsf{g}\} \equiv \frac{\partial \mathsf{f}}{\partial \cos \theta} 
\frac{\partial \mathsf{g}}{\partial \phi} - \frac{\partial \mathsf{f}}{\partial \phi} 
\frac{\partial \mathsf{g}}{\partial \cos \theta}, \quad 
\forall ~ \mathsf{f},~\mathsf{g}~\text{defined on the sphere}\label{fgbrac}
\eea
where $\theta = [0,\pi]$ and $\phi=[0,2\pi]$ are angular coordinates and $\{\mathsf{f},~\mathsf{g}\}$ is 
the Poisson bracket of continuous functions $\mathsf{f}$ and $\mathsf{g}$ on the sphere. We notice that 
although generators $\hL_{lm}$ are linear combination of $\partial / \partial (\cos \theta)$ and 
$\partial / \partial \phi$, the~latter operators cannot be considered as generators of 
$SU(N \rightarrow \infty)$, because they commute with each other and generate only the abelian 
subspace of $SU(\infty)$ group.

Using (\ref{lharminicexp})--(\ref{fgbrac}), coefficients $f^{l"m"}_{lm,l'm'}$ can be determined:
\be
f ^{l"m"}_{lm,l'm'} = \frac{(2l" + 1)}{4\pi} \int d^2\Omega~ Y^*_{l"m"} \{Y_{lm},~Y_{l'm'}\}, \quad\quad 
Y^*_{lm} = Y_{l,-m} \quad \quad d^2\Omega \equiv d(\cos \theta)~d\phi  \label{flm}
\ee

Here we normalize $Y_{lm}$ such that:
\be
\int d^2\Omega ~ Y^*_{l'm'}~Y_{lm} = \frac{4\pi}{(2l+1)} \delta_{ll'}~\delta_{mm'} \label{ynorm}
\ee

Although $\hL_{lm}$ is defined in discrete $(l,m)$ space---analogous to a discrete Fourier mode---we can use inverse expansion to define operators which depend only on continuous angular coordinates:
\be
\hL ({\theta, \phi}) \equiv \sum_{l,m} Y^*_{lm} \hL_{lm} \label{xdef}
\ee

As $\{\hL (\theta, \phi)\}$ are linear in $\hL_{lm}$ and contain all these generators, they are also generators 
of $SU(N\rightarrow \infty) \cong \text{Diff} (S_2)$ and coefficients in their Lie bracket is expressed 
with respect to $\theta$ and $\phi$ as:
\be
\cf((\theta,\phi),(\theta',\phi'); (\theta",\phi")) = \sum_{lm,l'm',l"m"} Y^*_{lm}(\theta,\phi) 
Y^*_{l'm'}(\theta',\phi') Y_{l"m"}(\theta",\phi'') f ^{l"m"}_{lm,l'm'} \label{fthetaphidef}
\ee

Coefficients $\cf$ are anti-symmetric with respect to the first two sets of parameters and can be 
considered as a 2-form on the sphere, and Lie algebra of $\hL (\theta, \phi)$ operators as:
\be
[\hL (\theta_1,\phi_1), \hL (\theta_2,\phi_2)] = \int d\Omega_3 ~ \cf((\theta_1,\phi_1),(\theta_2,\phi_2); 
(\theta_3,\phi_3)) ~ \hL (\theta_3,\phi_3) \label{xcommute}
\ee

Operators $\hL (\theta, \phi)$ are continuous limits of $\hL_{lm}$'s and both set of 
generators are vectors and live on the tangent space of the~sphere.

\section {Cartan Decomposition of \boldmath{$SU(\infty)$}} \label{app:cartandecomp}
Representations of $su(N)$ algebra can be decomposed to direct sum of smaller $su$ algebras, see~e.g.,~\cite{cartandecomp} and references therein. In~the case of $SU(\infty)$ the fact that its algebra is 
homomorphic to Poisson brackets of spherical harmonic functions, which in turn correspond to 
representations of $SU(2) \cong SO(3)$, means that $su(\infty)$ algebra should be expandable as direct 
sum of representations of $SU(2)$, see e.g.,~\cite{suninfhoppthesis,suninfym} for the proof. Thus, up~to a 
normalization factor depending only on $l$, generators of $su (\infty)$ algebra $\hL_{lm}$ can be expanded as:
\be
\hL_{lm} = \mathcal {R} \sum_{i_\alpha = {1,~2,~3}, \alpha = {1, \cdots, l}} a^{(m)}_{i_1, \cdots i_l} \sigma_{i_1} \cdots 
\sigma_{i_l} \label {llmdef}
\ee
where $\sigma_{i_\alpha}$'s are $N \rightarrow \infty$ representation of Pauli 
matrices~\cite{suninfhoppthesis}. Coefficients $a^{(m)}$ are determined from expansion of spherical 
harmonic functions with respect to spherical description of Cartesian coordinates, 
see~\cite{suninfhoppthesis} for details. This explicit description shows that up to a constant factor $\hL_{lm}$ 
operators can be considered as tensor product of $2 \times 2$ Pauli matrices, and~
$SU(\infty) \cong SU(2) \otimes S(2) \otimes \ldots$. This relation can be understood from properties 
of $SU(N)$ group. Specifically, $SU(N) \supseteq SU(N-K) \otimes SU(K)$. For~$N \rightarrow \infty$ and 
finite $K$, $SU(N-K \rightarrow \infty) \cong SU(\infty)$. Therefore $SU(\infty)$ is homomorphic to 
infinite tensor product of $SU$ groups of finite rank, in~particular $SU(2)$---the smallest non-abelian 
$SU$ group. This shows that $SU(2)$ group, which has a key role in some quantum gravity models, notably 
in LQG, simply presents a mathematical description rather than a fundamental physical entity. 
The description of $SU(\infty)$ as tensor product of $SU(2)$ is comparable with Fourier transform, 
which presents the simplest decomposition to orthogonal functions, but~can be replaced by another orthogonal 
function. It is only the application that determines which one is more~suitable.

\subsection {Eigen Functions of $\hL (\theta, \phi)$ and $\hL_{lm}$} \label{app:eigen}
We define eigen functions of $\hL (\theta, \phi)$ and $\hL_{lm}$ operators as the followings:
\bea
\hL (\theta, \phi) \eta (\theta, \phi) & = & N \eta (\theta, \phi) \label {xeigen} \\
\hL_{lm} \zeta_{lm} & = & N' \zeta_{lm} \label{leigen}
\eea
where $N$ and $N'$ are constants\footnote{A priori $N$ and $N'$ can depend on $(\theta, \phi)$. However, their 
dependence on angular parameters can be included in $\eta$. Therefore, only constant eigen values matter.}, but 
$N'$ may depend on $(l,m)$. Using definition of $\hL (\theta, \phi)$ and $\hL_{lm}$ and properties of spherical 
harmonic functions, solutions of Equations~(\ref{xeigen}) and (\ref{leigen}) are obtained as:
\be
\begin{cases}
\eta (\theta, \phi) = iN \sum \limits_{lm} ~ \frac{(l+m)!}{m A_l (l-m)!} ~  
[\mathcal{F}_{lm} (\cos \theta) - \mathcal{F}_{lm} (\cos \theta_0 (t))] + \eta (\theta_0 (t)) &  \\ 
\phi + H (\cos \theta) = -[H (\cos \theta_0(t)) - \phi_0 (t)] & \\
\end{cases} \label{etaparamsol} \\
\ee
\bea
&& A_l \equiv \sqrt{\frac{4\pi}{2l+1}} \quad \quad \mathcal{F}_{lm} \equiv \int d(\cos \theta) 
|P_{lm} (\cos \theta)|^{-2}, \label{alfdef} \\
&& H (\cos\theta) \equiv  \int d(\cos \theta) \frac{\sum \limits_{lm} \frac{A_{l} (l-m)!}{(l+m)!} ~ 
\frac{\partial |P_{lm} \cos \theta|^2}{2 \partial \cos \theta}}
{\sum \limits_{l'm'} {\frac{im' A_{l'} (l'-m')!}{(l'+m')!} ~ |P_{l'm'} (\cos \theta)|^2}} \label{fhdef}
\eea
where $t$ parameterizes tangent surface at initial point $(\theta_0, \phi_0)$. Elimination of this parameter 
from two equations in (\ref{etaparamsol}) determines $\eta (\theta,\phi)$ for a set of initial conditions. 
Because the second equation does not depend on $N$, without~loss of generality we can scale 
initial value $\eta (\theta_0) \rightarrow iN \eta (\theta_0)$. With~this choice the eigen value $N$ can be 
factorized, and~because Hilbert space is projective, $N$ can be considered as an overall normalization 
factor and irrelevant for physics. Therefore, each set of parameters $(\theta, \phi)$ present a unique 
pointer state for the Hilbert~space.

In the same way we can calculate eigen functions of $\hL_{lm}$ as a parametric function:
\be
\begin{cases}
\zeta_{lm} (\theta) = -N' e^{m^2 W_{lm} (\theta_0)} \sqrt{\frac{(l+m)!}{(l-m)!}} 
[Z_{lm} (\theta) - Z_{lm} (\theta_0 (t))] + \zeta_{lm} (\theta_0 (t)) & \\
\phi - im W_{lm} (\theta) = \phi_0 (t) - im W_{lm} (\theta_0 (t)) \\
\end{cases} \label{zetaparamsol}  \\
\ee
\bea
&& W_{lm} \equiv \int d(\cos \theta) \frac {(1 - \cos \theta^2)~P_{lm} (\cos \theta)}{(l - m + 1)~
P_{(l+1)m} (\cos \theta)} - (l+1) P_{lm} (\cos \theta) \label{wdef} \\
&& Z_{lm} \equiv \int d(\cos \theta) \frac {e^{-m^2~W_{lm} (\cos \theta)}}{(l - m + 1)~
P_{(l+1)m} (\cos \theta)} - (l+1) P_{lm} (\cos \theta) \label{zdef}
\eea
Similar to $\eta (\theta,\phi)$, redefinition of initial value 
$Z_{lm} (\theta_0 (t)) \rightarrow Z_{lm} (\theta_0 (t)) N' e^{m^2 W_{lm} (\theta_0)}$ 
leads to a unique eigen function for $\hL_{lm}$.

Considering diffeomorphism invariance of the model, it is always possible to redefine coordinates such that 
$\theta = const.$ and $\phi = const.$ constitute a basis and any state can be written as:
\be
|\psi \rangle = \int d^2\Omega ~ \psi (\theta, \phi) |\theta, \phi \rangle \label{statedecomp}
\ee
Thus, as~explained in the main text, vectors of the Hilbert space representing $SU(\infty)$ are complex 
functions on 2D surfaces. As~$SU(\infty) \cong \text{Diff}(S_2)$, the~range of 
$(\theta, \phi)$ is $\theta = [0., \pi]$ and $\phi = [0., 2\pi)$. However, $SU(\infty)$ may be represented 
by diffeo-surfaces of higher genus. In~this case $|\theta + n\pi, \phi + 2n'\pi \rangle$ for 
any integer $n$ and $n'$ may present different states. States can be also expanded with respect 
to $|l,m\rangle$~\cite{suninfhoppthesis}. 

\subsection{Dynamics Equations of the Universe before its Division to~Subsystems} \label{app:dyneq}
The equilibrium solution for Lagrangian $\mathcal{L}_U$ in (\ref{twodlagrang}) can be determined by 
variation with respect to $L_{lm}$ and components of the state $\rho$ in an orthogonal basis of the 
Hilbert space. In~absence of environment for the whole Universe, $\rho$ is pure and can be written as 
$\rho = |\psi\rangle \langle \psi |$, where $\psi\rangle$ is an arbitrary vector in the Hilbert space 
$\hm_U$. As~discussed in Section~\ref{sec:infuniverse} and explicitly shown in Appendix \ref{app:eigen}, 
vectors of $\hm_U$ correspond to complex functions of angular coordinates $(\theta,\phi)$ of the 
diffeo-surface, and can be expanded with respect to spherical harmonic functions. Nonetheless, here we 
follow the usual bracket notation of quantum mechanics and call states of this 
orthogonal basis $|l,m\rangle$, where $l \in \mathbb {Z} /2,~-l \leqslant m \leqslant +l $. In~this basis 
$|\psi\rangle  =\sum_{l,m} \psi_{lm}~|l,m \rangle$ and 
$\rho = \sum_{l,m,l',m'} \psi_{l,m} \psi^*_{l',m'} |l,m \rangle \langle l',m'|$. After~this 
decomposition dynamics equations are expressed as:
\bea
\frac{\partial \mathcal{L}_U}{\partial \psi_{lm}} & = & \sum_{l',m',l",m"} L_{l"m"} \psi^*_{l'm'} \langle l'm'| 
\hL_{l"m"} |l,m \rangle \label{dypsi} \\
\frac{\partial \mathcal{L}_U}{\partial L_{lm}} & = & \sum_{l',m',l",m"} \psi^*_{l"m"} \psi_{l"m"} \langle l',m'| 
\hL_{lm} |l"m" \rangle + 2 L_{lm} \tr (\hL_{lm} \hL_{lm}) \label {dynl}
\eea

Because $\hL_{lm}$ is a generator of $SU(\infty)$, the~last term in (\ref{dynl}) is a constant depending only 
on $l$ and normalization of generators. Thus, we define $C_l \equiv \tr (\hL_{lm} \hL_{lm})$. Using 
description of $\hL_{lm}$ in (\ref{llmdef}) to tensor product of Pauli matrices, we conclude that 
$\hL_{lm} |l',m' \rangle \neq 0$ only for $l \geqslant l'$ and consists of linear combination of 
$|l'',m''\rangle$ states. On~the other hand, $\langle l',m' | l, m \rangle = \delta_{ll'} \delta_{mm'}$. 
Thus, $\langle l',m'| \hL_{lm} |l"m" \rangle$ is nonzero only for terms with equal $l$ indices and we can 
solve (\ref{dynl}) for $L_{lm}$ as the following:
\be
L_{lm} = -\frac{1}{2 C_l} \sum_{|m'|,|m''| \leqslant l, m+m'+m'' = 0} \psi^*_{lm'} \psi_{lm''} 
\langle l,m'|\hL_{lm} |lm" \rangle \label{llmsol}
\ee

By applying this solution to (\ref{dypsi}) and using properties of $\hL_{lm}$ and $|l,m \rangle$ we find:
\bea
&& \sum_{m'',m''} \psi^*_{l, -(m+m')}  \psi^*_{l, -(m+m'')} \psi_{lm'} \langle l, -(m+m'') | \hL_{lm''} | l,m \rangle 
 \langle l, -(m'+m'') | \hL_{lm''} | l,m' \rangle = 0. \nonumber \\
&& \quad \quad |m'|, |m'|, |m+m''|, |m'+m''|, \leqslant l, \forall l \in \mathbb {Z} /2 \label{psieq}
\eea

Considering independence and orthogonality of $|l,m \rangle$ states, this equation is satisfied only if 
\mbox{$\psi_{lm} = 0,~|m| \leqslant l,~\forall l$}. Thus, equilibrium solution of the Lagrangian 
$\mathcal{L}_U$ is a trivial~vacuum. 


\end{document}